\documentclass[fleqn,usenatbib]{mnras}

\usepackage{newtxtext,newtxmath}
\usepackage[T1]{fontenc}
\DeclareRobustCommand{\VAN}[3]{#2}
\let\VANthebibliography\thebibliography
\def\thebibliography{\DeclareRobustCommand{\VAN}[3]{##3}\VANthebibliography}
\usepackage{ae,aecompl}
\usepackage{placeins}

\usepackage[utf8]{inputenc}
\usepackage{graphicx}
\usepackage{color}
\usepackage{amsmath}
\usepackage{datetime}
\usepackage{hyperref}
\usepackage{subcaption}
\usepackage{booktabs}

\def\mnras{MNRAS}
\def\apj{ApJ}

\def\apjl{ApJL}
\def\nat{Nature}

\def\msun{\,\rm M_{\odot}}

\newcommand{\med}[1]{\left<#1\right>}

\title[Galactic feedback and dark-matter halo shapes]{The impact of galactic feedback on the shapes of dark-matter haloes}
\author[K.T.E. Chua et al.]{Kun Ting Eddie Chua$^{1}$\thanks{Email: \url{eddie@cheleb.com}}, Mark Vogelsberger$^2$, Annalisa Pillepich$^3$ and Lars Hernquist$^4$\\
$^{1}$Institute of High Performance Computing, 1 Fusionopolis Way, Singapore 138632\\
$^{2}$Department of Physics, Massachusetts Institute of Technology, 77 Massachusetts Avenue, Cambridge, MA 02139, USA\\
$^{3}$Max-Planck-Institut f{\"u}r Astronomie, K{\"o}nigstuhl 17, 69117 Heidelberg, Germany\\
$^{4}$Harvard-Smithsonian Center for Astrophysics, 60 Garden Street, Cambridge, MA 02138\\
}

\date{Accepted XXX. Received YYY; in original form ZZZ}
\pubyear{2021}

\begin{document}
	
\maketitle

\begin{abstract}

We quantify the impact of galaxy formation on dark matter halo shapes using cosmological simulations at redshift $z=0$.
The haloes are drawn from the IllustrisTNG project, a suite of magneto-hydrodynamic simulations of galaxies.
We focus on haloes of mass $10^{10-14} \msun$ from the 50~Mpc (TNG50) and 100~Mpc  (TNG100) boxes, and compare them to dark matter-only (DMO) analogues and other simulations e.g. NIHAO and Eagle.
We further quantify the prediction uncertainty by varying the baryonic feedback models in a series of smaller 25~${\rm Mpc}\,h^{-1}$ boxes.
We find that:
(i) galaxy formation results in rounder haloes compared to the DMO simulations, in qualitative agreement with past hydrodynamic models.
Haloes of mass~${\approx}2\times 10^{12} \msun$ are most spherical, with an average minor-to-major axis ratio of $\left< s \right> \approx 0.75$ in the inner halo, an increase of 40 per cent compared to their DMO counterparts.
No significant change in halo shape is found for low-mass $10^{10} \msun$ haloes;
(ii) stronger feedback, e.g. increasing galactic wind speed, reduces the impact of baryons;
(iii) the inner halo shape correlates with the stellar mass fraction, which can explain the dependence of halo shapes on different feedback models;
(iv) the fiducial and weaker feedback models are most consistent with observational estimates of the Milky Way halo shape. 
Yet, at fixed halo mass, very diverse and possibly unrealistic feedback models all predict inner halo shapes that are closer to one another than to the DMO results.
This implies that a larger observational sample would be required to statistically distinguish between different baryonic prescriptions due to the large halo-to-halo variation in halo shapes.

\end{abstract}


\section{Introduction}
\label{sec:intro}

The rise in computational power has led to an increase of the scope of hydrodynamic cosmological simulations. 
By implementing galaxy formation physics, these simulations aim to model the formation of realistic galaxies in a cosmological context \citep[for a review, see][]{Vogelsberger20v2}.
In addition to their effects on the stellar and gaseous content of galaxies,
baryonic processes such as radiative cooling, star formation as well as stellar and AGN feedback can also have significant effects on the structure of their host dark matter haloes and subhaloes.

In particular, theoretical predictions for halo shapes have been studied extensively using cosmological simulations.
Hierarchical structure formation predicts anisotropic halo growth, since the accretion of matter onto dark matter (DM) haloes during their growth is preferentially aligned with dark matter sheets and filaments.
This prediction is well-supported by a wealth of $N$-body simulations \citep[e.g.][]{Dubinski91v378,Warren92v399,Bullock2002,Jing02v574,Bailin05v627,Allgood06v367,Maccio08v391,Vera-Ciro2011}, which have shown that dark matter-only (DMO) haloes are triaxial and prolate ($c/b > b/a$)\footnote{We denote the major, intermediate, and minor axes semi-lengths as $a,b$ and $c$, respectively, with $a>b>c$.}.
In Milky Way (MW)-sized haloes (${\sim} 10^{12} \msun$), DMO simulations predict a sphericity $c/a \approx 0.5-0.6$ near the galactic centre (within few tens of kpc).
These results are, however, in contrast to the Milky Way (MW) halo shape estimated using stellar kinematics and stellar streams, which suggest a more spherical inner halo e.g. $c/a \ge 0.8$ \citep{Ibata01v551}, and $c/a = 0.72$ \citep{Law10v714}.

The condensation of baryons to halo centres can have an impact on halo shapes, since the formation of a central baryonic mass can scatter and modify the orbits of approaching dark matter particles.
Such an effect was derived in \cite{Debattista08v681}, which simulated an isolated halo and found the growth of a central component can deform box orbits into rounder trajectories.
Consequently, haloes from galaxy formation simulations have been found to be transformed into more spherical configurations than their analogues from DMO simulations \citep[e.g.][]{Katz91v377,Katz93v412,Dubinski94v431,Tissera10v406,Abadi10v407,Kazantzidis10v720,Bryan13v429,Butsky16v462,Chisari2017,Chua19v484}.
This has also led to better agreement between halo shapes in baryonic simulations and the observationally-inferred dark matter halo shape  of our Galaxy \citep[e.g.][]{Chua19v484, Prada19v490}.
More recently, \cite{Cataldi21v501} further concluded using the Fenix and {\sc eagle} simulations that dark matter halo shapes are related to the morphology of the central galaxies, pointing to a fundamental relation between the growth of the central galaxy and its parent halo.

Despite the role of galaxy formation models in resolving disagreements between $N$-body predictions and observations, as well as in reproducing stellar galactic properties, significant uncertainty regarding feedback implementations in simulations remains.
Such variations arise because many baryonic processes such as star formation and black hole accretion are not directly resolved in most galaxy formation simulations.
Instead, these effects are described through sub-grid models, where  multiple parametrizations often exist.
In many cases, large galaxy formation simulations rely on a single fiducial galaxy formation model, where parameters and parametrizations are tuned or chosen to reproduce a set of observations.
As such, many numerical studies relying on these simulations do not take into account inherent uncertainty in baryonic feedback, which can have important consequences on the predicted properties of the galaxy and its surrounding medium \citep[e.g.][]{Suresh15v448}.

In the absence of outflows and star formation, the gaseous disc that forms at the centres of simulated  haloes is unrealistically small and massive, but provides an idea of the maximal effect of galaxy assembly on the dark matter haloes.
For example, such a configuration was investigated by \cite{Abadi10v407}, who estimated such haloes to be completely oblate with axial ratios of the iso-potential contours to be $b/a \sim 1$ and $c/a \sim 0.85$.
Although newer cosmological simulations allow realistic galaxies to form in a cosmological context and enable a more accurate characterization of halo shapes \citep[e.g.][]{Butsky16v462,Chua19v484,Prada19v490, Emami2021v913}, these studies often consider only one `fiducial' galaxy formation model.
One source of uncertainty in predictions of halo shapes from galaxy formation simulations can arise from the strength of feedback prescriptions. 
Using the OWLs simulations \citep{Schaye10v402}, \cite{Bryan13v429} showed that varying the stellar and AGN feedback can lead to notable changes in the dark matter halo shapes.
However, the non-iterative method that they used to calculate halo shapes was not able to account for the variation of halo shapes with radius and has been shown to be less accurate compared to methods based on an iterative procedure \citep{Zemp11v197}.

In this paper, we focus on understanding how the shapes of dark-matter haloes are affected by variation in galactic feedback.
We begin by analysing haloes from IllustrisTNG\footnote{\url{www.tng-project.org}} \citep[hereafter, TNG:][]{Marinacci18v480, Naiman18v477, Nelson17v475, Pillepich18v475, Springel18v475}, a suite of cosmological magneto-hydrodynamical (MHD) simulations  carried out in various box sizes. 
The IllustrisTNG simulations were conceived to improve upon the original Illustris simulation \citep{Vogelsberger13v436, Vogelsberger14v444,Illustris,Genel14v445}, which was found to exhibit tensions with observations in terms of:
1) a cosmic star formation rate density at $z<1$ that was too high;
2) a stellar mass function and stellar mass fraction at $z=0$ that were too high for both high-mass and low-mass galaxies,
3) $z=0$ galaxy sizes (stellar half-mass radii) that were too large, and 4) halo X-ray emission from high-mass galaxies severely underestimating that seen in observations.
The updated TNG galaxy formation model alleviates the above-mentioned tensions while continuing to demonstrate good agreement with observational constraints \citep[see][for a partial list]{Nelson19TNGrelease}.
Through radiative transfer post-processing, the suite of IllustrisTNG simulations has  also provided detailed predictions of the high redshift galaxy populations expected to be observed by the James Webb Space Telescope \citep{Vogelsberger20v492, Shen20v495, Shen2021}.
In terms of halo shape, the similarity in box size, element count and initial conditions facilitates a direct comparison  between the 100~Mpc box (TNG100) and the older Illustris, which we had previously analysed in \cite{Chua19v484}.

In addition to flagship IllustrisTNG runs, we analyse an additional set of simulations that include variations of the TNG feedback model \citep[see][for more details]{Pillepich18v473}.
While these simulations are carried out in smaller boxes of side length $25h^{-1}$~Mpc, they provide important insights into the relationship between the feedback strength and halo shapes.
By making specific variations to feedback parameters of the galactic wind and AGN model, we can further and systematically quantify the impact of baryonic physics on dark matter halo shapes.

The paper is structured as follows:
we describe our simulation methods and definitions in Section~\ref{sec:methods}.
We present the main results on halo shapes from TNG100 and TNG50 in Section~\ref{sec:results}, and compare them to those from other hydrodynamic simulations.
In Section~\ref{sec:variations}, we analyse the impact of feedback variations using the smaller boxes, and present a comparison to observational estimates of the Milky Way halo shape. 
Finally, our summary and conclusions are presented in Section~\ref{sec:conclusions}.

\section{Methods and Definitions}
\label{sec:methods}

\subsection{IllustrisTNG Simulations}

\begin{table*}
\begin{tabular*}{\textwidth}{@{\extracolsep{\fill}}l c c c  c c c c}
\hline
Name	& Type & Box size	& DM particles \& cells	&  $m_{\rm DM}$		& $m_{\rm baryon}$ & $\epsilon^{z=0}_{\rm DM,stars}$ \\ 
& 	& [${\rm Mpc}$] & &  [$10^6 \msun$]	& [$10^6 \msun$]  & [kpc]	\\
\hline
TNG50 & MHD & 35$h^{-1}\approx 52$		& $2\times 2160^3$ &  0.454 	& 0.085 & 0.288 \\
TNG50-DM & DMO & 35$h^{-1}$		& $2160^3$ &  0.54 	& - & 0.288 \\ 
TNG100 & MHD & 75$h^{-1} \approx 110$		& $2\times 1820^3$ &  7.5 	& 1.4  & 0.74 \\ 
\rule{0pt}{4ex}TNG100-DM & DMO	& 75$h^{-1}$ & $1820^3$	& 8.9 & - & 0.74  \\ 

\rule{0pt}{4ex}Illustris & Hydrodynamic	& 75$h^{-1}$ & $2\times 1820^3$ & 6.3 & 1.3  & 1.42/0.71 \\ 
Illustris-Dark & DMO	& 75$h^{-1}$ & $1820^3$ &  7.52	& - & 1.42 /-\\ 

\rule{0pt}{4ex}L25n512 & Various & 25$h^{-1} \approx 37$ & $2\times512^3$ & 12.4 & 2.4 &  0.74\\
               (Small boxes) & (see Table \ref{table:variations}) & &  &  & &  \\
\hline
\end{tabular*}
\caption{Summary of the main simulations and their resolution parameters examined in this work:
	(1) simulation name;
	(2) the type of simulation;
	(3) length of simulation box;
	(4) number of cells and particles in the simulation;
	(5) mass per DM particle;
	(6) target mass of baryonic particles;
	(7) Plummer-equivalent gravitational softening lengths at redshift $z=0$.
In the TNG runs, the softening lengths for all particle types are co-moving kpc for $z>1$, after which they are fixed to their $z=1$ values in physical space.
Note that in Illustris, this procedure is not applied to the DM particles, thus DM particles have twice the softening lengths (first value) as the stellar particles (second value).
The MHD simulations are based on the fiducial TNG galaxy formation model described in \protect\cite{Pillepich18v473}.
The nine model variations in the small boxes are further listed in Table \ref{table:variations}. 
}
\label{table:parameters}
\end{table*}

The haloes we study in this work are drawn from \emph{The Next Generation Illustris Simulations} (IllustrisTNG), a suite of cosmological magneto-hydrodynamical simulations \citep{Marinacci18v480, Naiman18v477, Nelson17v475, Pillepich18v475, Springel18v475}.
The simulations are performed using the simulation code {\sc Arepo} \citep{Springel09v401}, which calculates the gravitational forces using a Tree-Particle-Mesh method and solves the ideal magneto-hydrodynamic (MHD) equations using a finite volume method on an adaptive mesh.
The cosmology utilized in IllustrisTNG is consistent with that of Planck, given by $\Omega_m = 0.3089$, $\Omega_\Lambda = 0.6911$, $\Omega_b = 0.0486$, $\sigma_8 = 0.8159$, $n_s = 0.9667$ and $h = 0.6774$ \citep{Planck,Spergel15v91}.

The galaxy formation model adopted in TNG accounts for 
(i) primordial and metal-line gas cooling, 
(ii) a spatially-uniform and time-dependent UV background, 
(iii) stellar formation and  feedback, and 
(iv) kinetic and thermal feedback from black holes.
The TNG galaxy formation model builds upon the physics model introduced previously in  Illustris to address the shortcomings which were identified.
Compared to Illustris, these improvements include
(i) an updated kinetic AGN feedback model at low accretion rates, 
(ii) improved isotropic galactic winds, and
(iii) ideal magneto-hydrodynamics.
For specific details, we refer the reader to \cite{Weinberger17v465} and \cite{Pillepich18v473}.
The model parameters of IllustrisTNG are that of the default (\emph{fiducial}) model described in \cite{Pillepich18v473}, selected to produce good agreements between simulated galaxies and observations of the cosmic star formation rate density and the stellar content of the galaxy population at $z=0$.

In this project, we analyse halo shapes from both TNG50 \citep{Pillepich19v490, Nelson19v490} and TNG100, focusing primarily on the highest resolution simulation of each box size.
TNG50 has a volume of (50 Mpc)$^3$, with a dark matter mass resolution of $4.5 \times 10^5$ and baryonic mass resolution of $m_b = 8.5 \times 10^4 \msun$. 
TNG100 has a volume of (100 Mpc)$^3$, with a dark matter mass resolution of $7.5 \times 10^6$ and baryonic mass resolution of $m_b = 1.4 \times 10^6 \msun$. 
The lower resolution run TNG100-2 uses $2\times910^3$  elements with two times worse spatial resolution, while TNG100-3 uses $2 \times 455^3$ elements with four times worse spatial resolution compared to TNG100.
For each box size, DMO simulations are performed, providing baseline $N$-body results for comparison.
The important parameters of the simulations are summarized in Table~\ref{table:parameters}.
In general, both the MHD and hydrodynamics simulations will be referred to as the \emph{Full-Physics} runs.

\subsection{TNG model and variations}

\begin{table}
\begin{tabular*}{0.48\textwidth}{@{\extracolsep{\fill}}l c}
\hline
Variation type &  Details	\\ 
\hline
Fiducial & $\bar e_w=3.6, \kappa_w = 3.6$  \\
Strong winds & Doubled wind energy $(\bar e_w=7.2)$\\
Weak winds & Halved wind energy $(\bar e_w=1.8)$\\
Fast winds & Doubled wind speed $(\kappa_w = 14.8)$\\
Slow winds & Halved wind speed $(\kappa_w = 3.2)$\\
No winds & No galactic winds\\
No BH &  No black holes\\
No BH kinetic mode\ & Only BH quasar mode\\
Dark matter-only (DMO) & No baryons\\
\hline
\end{tabular*}
\caption{Variations of the galaxy formation model simulated in smaller boxes (L25n512). Each parameter or choice change is made with respect to the fiducial TNG model. Further description and usages of these model-variation runs can be found in \citealt{Pillepich18v473}.
}
\label{table:variations}
\end{table}

Besides the primary 50 Mpc and 100 Mpc boxes, variations of the fiducial TNG model were also simulated in smaller boxes of size $L = 25 h^{-1} \sim 37$~Mpc, with $2 \times 512^3$ resolution elements.
As shown in Table \ref{table:parameters}, these L25n512 small boxes have similar softening lengths and resolutions (within a factor of two) compared to TNG100.
All simulations of model variations were carried out with the same initial conditions, which were chosen by conducting low-resolution DMO simulations from 10 different initial density fields and then choosing the realization with the dark matter halo mass function closest to the average \citep{Pillepich18v473}.
This was done to reduce the effect of sample variance in this smaller box.

In this work, we focus on the effect of galactic winds and black holes.
In the TNG model, galactic winds are driven by star formation, which launches wind particles with an initial speed $v_w$ scaling with the local dark matter velocity dispersion $\sigma_\text{DM}$:
\begin{equation}
    v_w = \max \left[ \kappa_w \sigma_\text{DM} \left(\frac{H_0}{H(z)}\right) ^{1/3}, v_{w,\text{min}} \right]    
\end{equation}
where $H(z)$ is the redshift dependent Hubble parameter, and $v_{w,\text{min}}$ is a velocity floor.
$\kappa_w$ is thus a dimensionless factor that determines the speeds of launched winds, taken to be $\kappa_w=7.4$ in the fiducial model.
We investigate variations with doubled  $\kappa_w$ (\emph{Fast winds}) and with halved $\kappa_w$ (\emph{Slow winds}).

For a given wind speed, mass loading of the wind depends on the energy available for wind generation, which
is directly proportional to the free parameter $\bar e_w$ in the TNG model.
This quantity $\bar e_w$ reflects the energy released per core-collapse supernova, taken to be $\bar e_w=3.6$ in the fiducial case.
We examine the effects of increasing the wind energy, with this factor doubled to $\bar e_w=7.2$ (\emph{Strong winds}).
Additionally, a case without galactic winds (\emph{No winds}) is also carried out.

In the TNG model, black hole feedback is driven by a combination of thermal feedback heating the surrounding gas at high accretion rates, and a kinetic AGN feedback model driving black hole-driven winds at low accretion rates \citep{Weinberger17v465}.
We investigate the effect of suppressing kinetic AGN feedback (\emph{No BH kinetic winds}), which has been found to be important in quenching star formation in high-mass haloes.
Finally, a case neglecting black hole feedback (\emph{No BHs}) is also examined.

The nine model variations (including the fiducial and DMO models) are summarized in Table \ref{table:variations}.

\subsection{Identifying and matching haloes and subhaloes}

\begin{table*}
	\centering
	\begin{tabular*}{0.6\textwidth}{@{\extracolsep{\fill}}l c c c c c}
		\hline
		& \multicolumn{5}{c}{Halo mass $M_{200}\,[\msun]$}\\
		Simulation & $10^{10-11}$ & $10^{11-12}$ & $10^{12-13}$  & $10^{13-14}$ & $10^{14-15}$\\ 
		\hline
		TNG50       & 9179 & 1441   & 183  & (23)   &  (1)\\
		TNG50-DM    & 11398 & 1537  & 193   &  (24)   &  (1) \\
		TNG100      &  (87025) & 12963     & 1708  & 168 & 14 \\
		TNG100-DM   &  (109853) & 13912     & 1743  & 198 & 13 \\
		Illustris   &  (82249) & 12875     & 1317  & 109 & 10 \\
		Illustris-Dark &  (93569) & 11781  & 1440  & 160 & 11 \\
		L25n512 (Fiducial model) & (3274) & 496 & 77 & 9 & 0\\
		\hline
	\end{tabular*}
	\caption{Number of haloes at redshift $z=0$ adopted in this paper and extracted from TNG100, TNG50, Illustris, and their DMO counterparts, according to mass.
	In the 50-Mpc boxes (TNG50 and TNG50-DM), we focus on haloes of mass $10^{10} - 10^{13} \msun$.
	In the other simulations, which have lower resolutions, the minimum halo mass we consider is $10^{11} \msun$, hence the number of haloes in the boxes are given in parentheses. 
	For the smaller boxes (L25n512), the number of haloes corresponds to that of the fiducial model run.
	}
	\label{table:numhaloes}
\end{table*}

In the simulations, haloes are identified using a friends-of-friends ({\sc fof}) group finder algorithm with a linking length of 0.2\citep{Davis85}. 
The \textsc{subfind} algorithm subsequently identifies gravitationally self-bound subhaloes  \citep{Springel01v328,Dolag09v399}.
The subhalo in each {\sc fof} group with the lowest-potential resolution element is classified as {\it central}, and it is typically the most massive: the remaining subhaloes are called {\it satellites}, whether they include stars or not.
For each halo, the virial mass $M_{200}$ and virial radius $R_{200}$ are calculated\footnote{$R_{\Delta}$ is the radius within which the enclosed mass density is $\Delta$ times the critical value $\rho_c$, i.e. $\rho_{\rm halo} = \Delta \rho_c$. $M_{\Delta}$ is the total mass of the halo enclosed within $R_{\Delta}$. In this work, we use the value $\Delta = 200$.}.

In this work, we examine haloes at the current redshift $z=0$. 
Table \ref{table:numhaloes} compares the number of haloes in four mass bins in Full-Physics and DMO runs.
In TNG100, there are $\approx 14000$ haloes of mass greater than $10^{11} \msun$, including 14 cluster-sized ($\sim 10^{14}$) haloes.
Thus, we focus on haloes of small to intermediate masses ($10^{10} - 10^{13} \msun$) in TNG50, since its higher resolution allows the shapes of lower mass haloes to be resolved.

We also identify matching haloes between the Full-Physics and DMO runs to enable one-to-one halo shape comparisons between each halo pair.
For any given subhalo in the Full-Physics run, the matching DMO subhalo is the one with the largest fraction of matching dark matter particles, identified using their unique IDs.
The process is reversed, and only bidirectional matches are considered to be successful.
At the masses considered in this work, almost all central subhaloes are successfully matched:
This corresponds to 10804 out of 10827 haloes with $M_{200} \ge 10^{10} \msun$ in TNG50, and 14778 out of 14853 with $M_{200} \ge 10^{11} \msun$ in TNG100.

\subsection{Defining the halo shape}

The algorithm we used to quantify halo shapes is identical to that described in \cite{Chua19v484}, which we summarize as follows.
Assuming that dark matter haloes are triaxial, their shapes are determined by the axis ratios of the iso-density surface, $q\equiv b/a$ and $s\equiv c/a$ where $a$, $b$ and $c$ are the major, intermediate and minor axes respectively.
These parameters can be found using an iterative algorithm with the  {\it unweighted shape tensor} \citep[e.g.][]{Bailin05v627,Zemp11v197}:
\begin{equation}
	S_{ij} = \frac{1}{\sum_k m_k} \sum_k  m_k\, r_{k,i} \,r_{k,j},
	\label{eqn:shapetensor}
\end{equation}
related to the second moment of the mass distribution.
Here, $m_k$ is the mass of the $k$th particle, and $r_{k,i}$ is the $i$th component of its position vector.

The iterative algorithm allows the shape of the integration volume to adapt to the shape of the halo.
The first iteration begins with a spherical shell ($q=s=1$), within which a set of particles is selected. 
We then calculate and diagonalize the shape tensor using Eq. \ref{eqn:shapetensor}. 
The eigenvectors denote the directions of the principal axes, while the eigenvalues are related to the square-roots of the principal axes lengths ($i\propto\sqrt{\lambda_i}, \, i = a,b,c$).
In subsequent iterations, the new values of $q=b/a$ and $s=c/a$ are used to select a new set of particles, and the ellipsoidal shell is deformed keeping the semi-major length constant.
The process continues until both $q$ and $s$ converge; i.e. when fractional differences in successive values differ by less than 1 per cent.

To obtain the shape profiles, we calculate the local shape $q(r)$ and $s(r)$ in ellipsoidal shells as a function of distance from the halo centre.
The elliptical radius  
 \begin{equation}
 	r_{\rm ell}^2 = x^2 + \frac{y^2}{q^2} + \frac{z^2}{s^2}
 	\label{eqn:rell}
 \end{equation}
is used to determine if a given particle falls within an ellipsoidal shell with axis ratios $q$ and $s$.
Assuming that ellipsoids are oriented with $x$ along the  major axis and $z$ along the minor axis, 
$r_{\rm ell}$ is also the semi-major length of the  ellipsoid where the particle is residing.
Throughout this paper, halo-centric distances involving the halo shape will in general refer to this elliptical radius $r_{\rm ell}$.

When using shell-enclosed particles, the density distribution and thus the shape tensor can be sensitive to the presence of large satellites \citep{Zemp11v197}.
We avoid substructure contamination by considering only particles bound to the central subhalo, as identified by {\sc subfind}.
Using the unweighted shape tensor, we apply the iterative procedure to calculate the shape of each halo.
To obtain a radial profile, 15 ellipsoidal shells logarithmically-spaced between $0.01 \leq r/R_{200} \leq 1$ are used, each with a logarithmic width of $\Delta (r/R_{200}) = 0.1$~dex.
Finally, the triaxiality parameter $T\equiv(1-q^2)/(1-s^2)$ measures how prolate ($T=1$) or oblate ($T=0$) the halo is.
The axis ratios $q$ and $s$, as well as the triaxiality $T$ are collectively termed the halo shape parameters.


\section{Impact of baryons in TNG100 and TNG50}
\label{sec:results}

\begin{figure}
    \centering
    \includegraphics[width=\linewidth,trim={0.4cm 0.4cm 0.4cm 0.4cm},clip]{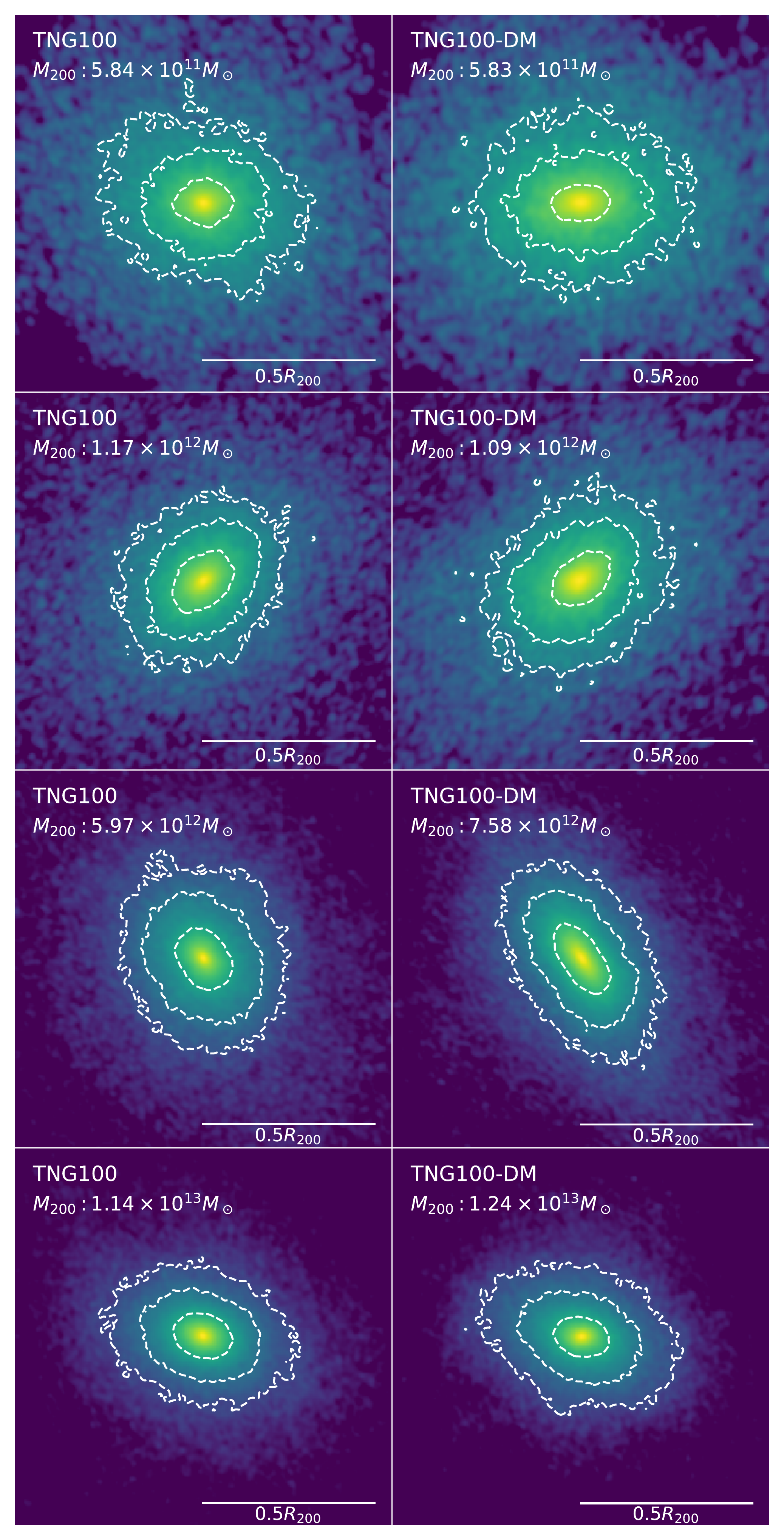}
    \caption{{\bf Dark matter density on a logarithmic colour scale within a slice of thickness 30~kpc, for TNG100 and TNG100-DM.}
    Four representative haloes in TNG100 (left) with their matched DMO counterparts in TNG100-DM (right) are shown, and the haloes are randomly projected on a 2D-plane.
    Dashed lines indicate iso-density contours.
    These examples highlight the sphericalization of haloes when including galaxy formation physics in the simulations.
    Additionally, we see that halo shapes are not uniform, but vary with distance to the halo centre.
    }
    \label{fig:tng100_visualization}
\end{figure}

\begin{figure}
    \centering
    \includegraphics[width=\linewidth,trim={0.4cm 0.4cm 0.4cm 0.4cm},clip]{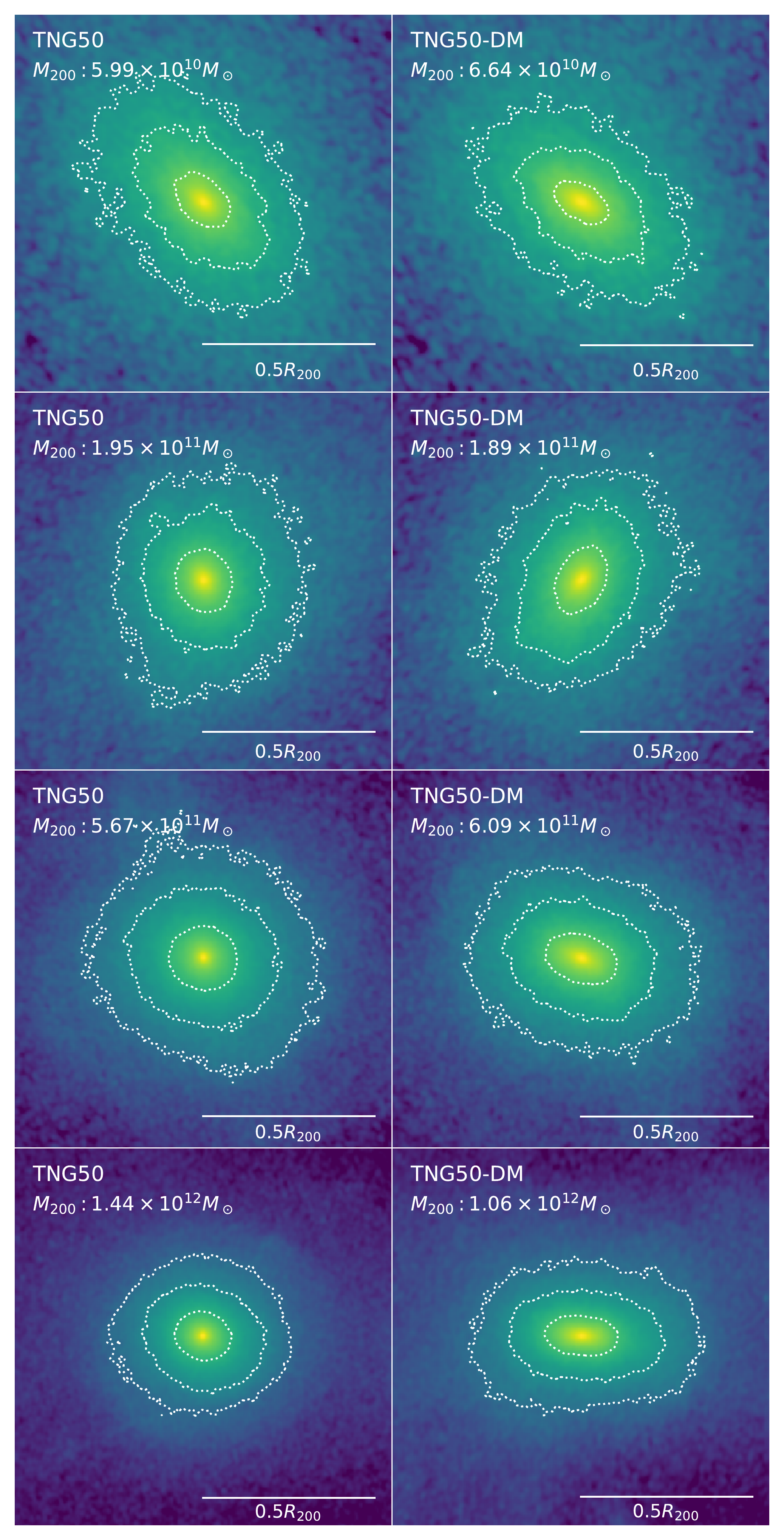}
    \caption{{\bf Dark matter density on a logarithmic colour scale within a slice of thickness 30~kpc, for TNG50 and TNG50-DM.}
    Compared to TNG100 (Figure~\ref{fig:tng100_visualization}), the density and contours are smoother due to the higher mass resolution in TNG50. 
    }
    \label{fig:tng50_visualization}
\end{figure}

We begin by visualizing halo shapes of some representative haloes of TNG100 and TNG50 in Figures~\ref{fig:tng100_visualization} and \ref{fig:tng50_visualization}, respectively.
Each plot shows the dark matter density in a 30~kpc-thick slice within half of the virial radius, for matching haloes in the MHD (left column) and DMO (right column) simulations.
To better highlight the halo shapes, contour lines corresponding to three different densities have also been included.
For each simulation, the haloes are ordered top to bottom, from less to more massive.

Qualitatively, it is apparent that the MHD haloes are more spherical than their DMO counterparts.
For example, the MHD haloes in the third and fourth rows of Figure~\ref{fig:tng50_visualization} (TNG50), appear almost completely spherical at these radii.
However, we caution making conclusions based on projections of the dark matter density since the projected shape can depend on the projection angle, and the small sample size can be unrepresentative of the halo population at large.

Although not a focus of this paper, we point out that apart from transforming halo shapes through the axis ratios $s$ and $q$ or the triaxiality $T$, the tilt of the halo can also be changed by baryonic physics.
This is most evident in the second row of Figure~\ref{fig:tng50_visualization}, where the major axis is aligned closer to the vertical in TNG50 compared to TNG50-DM, and can further impact the alignment of the galaxy with respect to the halo \cite[e.g.][]{Tenneti15v453, Velliscig15v454}.

\subsection{Determining resolved regions}

Due to limitations in the mass and spatial resolutions, not all regions of a halo can be reliably resolved in the simulations.
We determine the convergence radius $r_{\rm conv}$, the smallest radius for which a halo shape is resolved, by comparing the median DMO profiles at different resolutions (i.e. TNG100-DM with TNG100-DM-2 and TNG100-DM-3): see Appendix~\ref{sec:resolution}.
There we find that $\kappa(r_{\rm conv}) = 7$ describes the convergence radius well, where 
\begin{equation}
    \kappa(r) \equiv \frac{\sqrt{200}}{8} \frac{N(r)}{\ln N(r)} 
    \left[ \frac{\bar\rho(r)}{\rho_{\rm crit}} \right]^{-1/2}.
    \label{eqn:power}
\end{equation}
$N(r)$ is the number of dark-matter particles enclosed within a radius $r$, $\bar \rho(r)$ is the mean density within $r$, and $\rho_\text{crit}$ is the critical density of the universe.
The expression for $\kappa$ is derived from the ratio of the two-body relaxation timescale to the circular orbit timescale at the virial radius \citep{Power2003v338, Vera-Ciro2011}.

In TNG100-DM, the convergence radius is $r_{\rm conv} \approx 10$~kpc, which corresponds to ${\approx}$15 per cent of the virial radius for $10^{11} \msun$ haloes, and 5 per cent for $10^{12} \msun$ haloes.
Due to the higher mass and spatial resolutions in TNG50-DM, the convergence radius $r_{\rm conv} \approx 4$~kpc, corresponding to around 10 per cent of the virial radius in $10^{10} \msun$ haloes.
Further details of the convergence tests are presented in Appendix~\ref{sec:resolution}.
We note that $\kappa(r_{\rm conv}) = 7$ is consistent with high-resolution results from the Aquarius simulation, and allows the circular velocity to converge to better than 2.5 per cent \citep{Navarro10v402,Vera-Ciro2011}. 
It is not obvious how Eq.~\ref{eqn:power} should be applied or modified in the case of the Full-Physics runs, to account for their larger number and greater variety of resolution elements within the central regions of haloes. 
Although it is possible that the convergence radii of the halo shapes in e.g. TNG100 and TNG50 may be in principle smaller than those in TNG100-DM and TNG50-DM, we take the DMO estimates as reference throughout.


\begin{figure*}
    \centering
    \includegraphics[width=\textwidth]{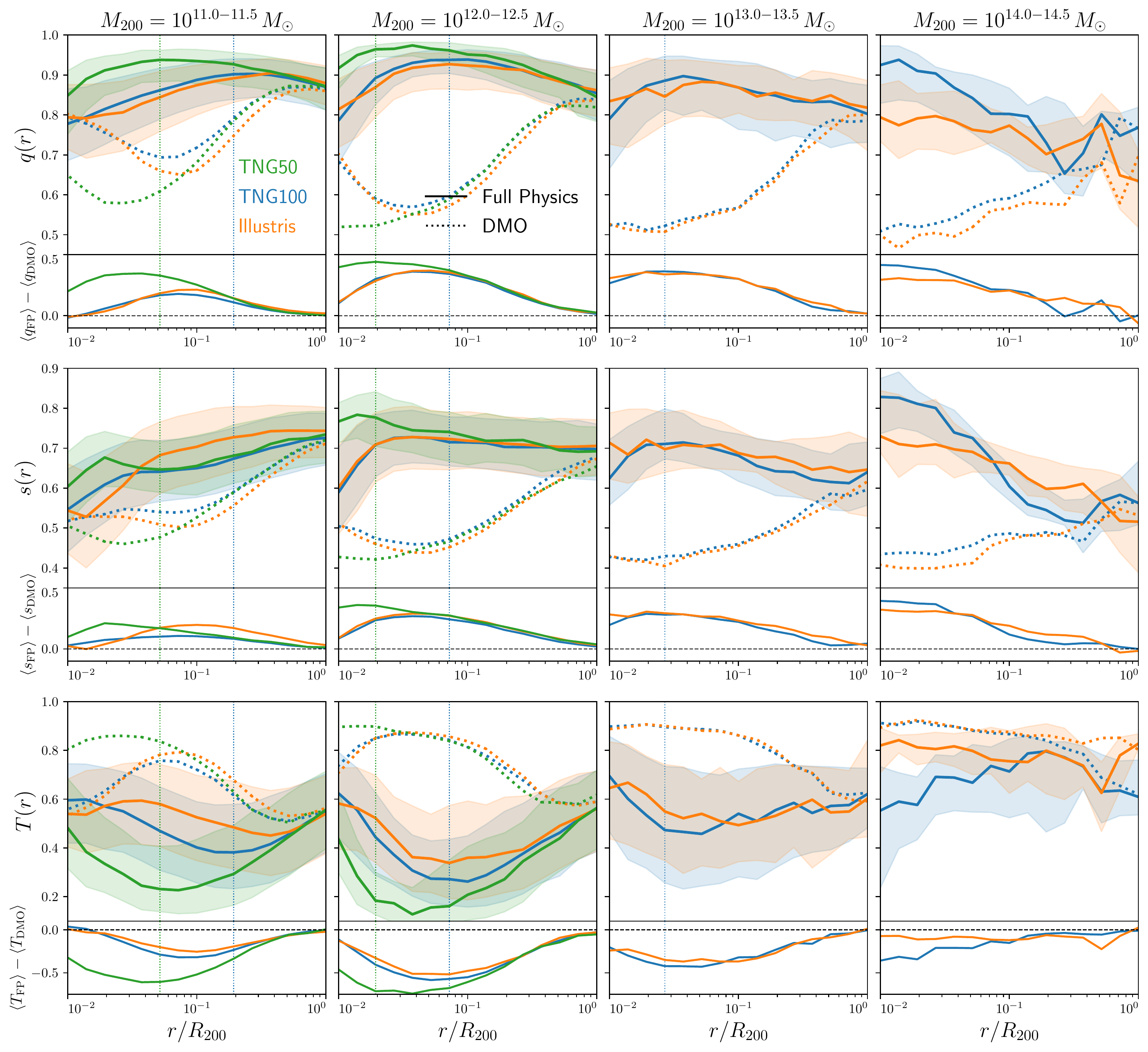}
    \caption{{\bf Shape profiles of dark matter haloes in TNG50 (green), TNG100 (blue) and Illustris (orange)}.
  	Haloes of mass between $10^{11}$ and $10^{14} \msun$ are selected in four mass intervals.
  	Solid curves denote the median profiles in the Full-Physics (MHD and Hydrodynamic) simulations, while dotted curves denote median profiles in the DMO simulations.
    The median convergence radius $r_{\rm conv}$ in TNG50 and TNG100 are indicated by the vertical green and blue lines, respectively.
    Halo-to-halo variations for the Full-Physics runs are represented by shaded regions which enclose the 25th-75th percentiles.
    The lower attached panels indicate the change in the shape parameters between the Full-Physics and DMO runs.
    In the Full-Physics simulations, the median halo becomes substantially more spherical (larger $q$ and $s$) and more oblate (smaller $T$) for all resolved radii, compared to the DMO results.
    From the lower attached panels, it is clear that the sphericalization is the largest near the halo centre, and diminishes towards the virial radius.
    In the DMO runs, the DMO haloes from TNG50-DM and TNG100-DM are also slightly more spherical and oblate on halo scales compared to Illustris-Dark: this is due to the larger $\sigma_8$ and $\Omega_m$ used in the TNG cosmology.
    }
    \label{fig:tng100_radial}
\end{figure*}

\begin{figure*}
    \centering
    \includegraphics[width=\textwidth]{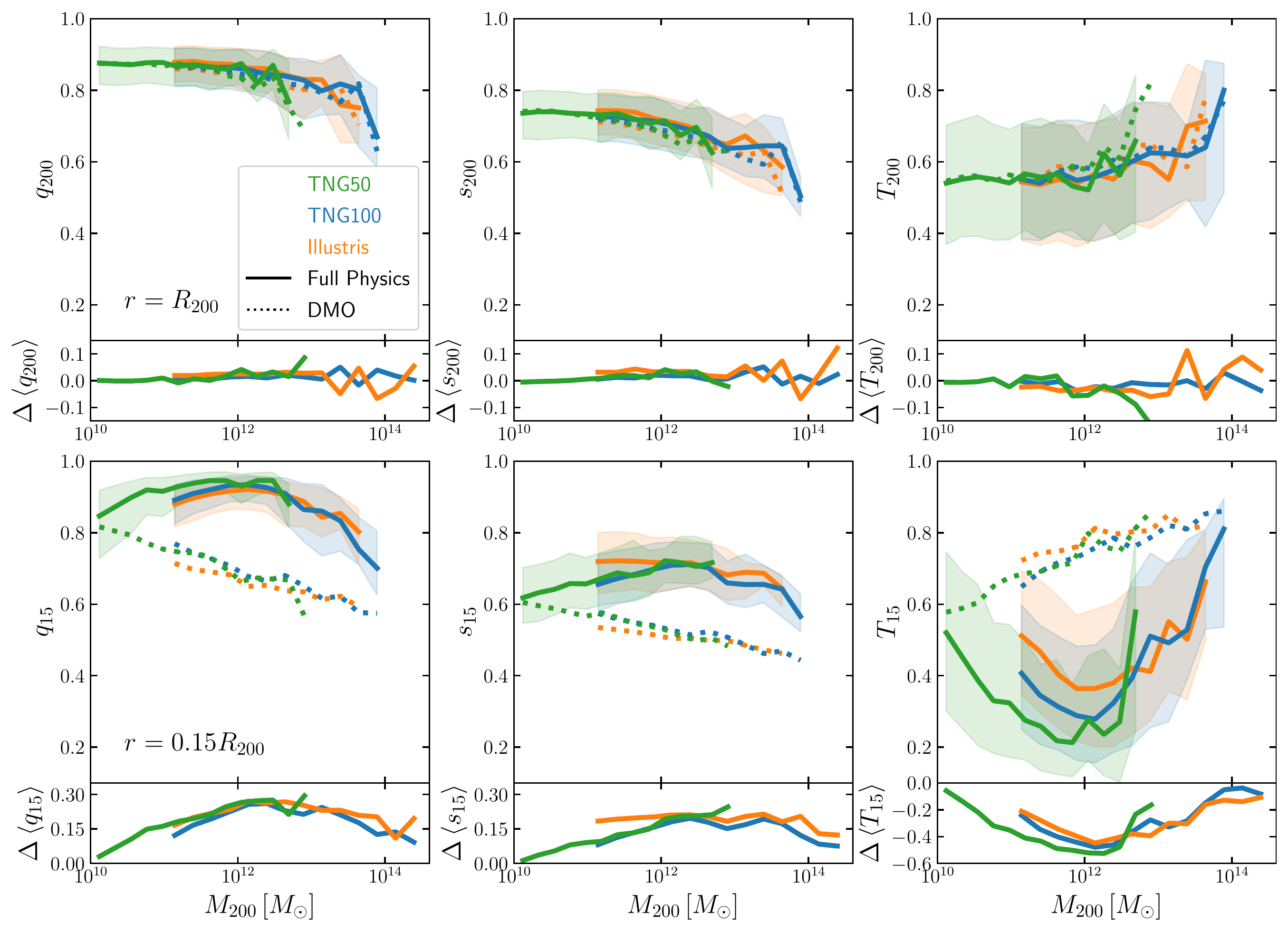}
    \caption{{\bf Dependence of dark matter halo shapes on halo mass at the virial radius (top panels) and in the inner halo (bottom panels, $r_{15} \equiv 0.15R_{200}$).}
    Solid lines denote results from the Full-Physics simulations (TNG and Illustris) while dotted lines denote results from the DMO simulations.
    The lower attached panels plot the difference in the median parameters between the Full-Physics and DMO simulations, i.e. $\Delta q = \left<q_{\rm FP}\right> - \left<q_{\rm DMO}\right>$.
    In TNG100, baryonic physics and the associated galaxy formation result in non-monotonic behaviour of the halo shapes, peaking at halo masses of $\msun \approx 2\times 10^{12} \msun$.
    }
    \label{fig:inner_m200}
\end{figure*}

\subsection{Radial profile of halo shapes in TNG}
\label{sec:radialprofile}

The radial profiles of the shape parameters $q$, $s$ and $T$ of TNG50, TNG100 and Illustris are shown in Figure~\ref{fig:tng100_radial} for haloes in different mass bins.
Solid lines show the median in each of the three Full-Physics (MHD and Hydrodynamic) simulations, while dotted lines correspond to the median results in the DMO counterparts (dotted lines).  
The shaded regions represent the 25-75th percentile of the distributions in the Full-Physics cases and illustrate the halo-to-halo variation of the halo shape.
The vertical dotted lines denote the convergence radii $r_{\rm conv}^{100}$ in TNG100-DM (blue) and $r_\text{conv}^{50}$ in TNG50-DM (green).

\subsubsection*{DMO results}

In the DMO simulations (dotted lines), dark matter haloes become more spherical and oblate towards the virial radius, a well-known result produced by a number of previous $N$-body studies \citep[e.g.][]{Allgood06v367,Hayashi07v377,Vera-Ciro2011}.
Although the shape profiles from all three DMO runs are in good agreement, we note that the average DMO halo from both TNG runs is slightly more spherical and more oblate compared to Illustris-Dark.
This effect is most evident for the lower mass $10^{11}$ and $10^{12} \msun$ haloes and is a result of the larger $\sigma_8$ and $\Omega_m$ parameters in the TNG cosmology ($\sigma_8=0.8159$, $\Omega_m=0.31$), compared to the  Illustris cosmology ($\sigma_8=0.81$, $\Omega_m=0.27$).
Thus, Illustris-Dark haloes form later and are less spherical on average \citep{Allgood06v367, Maccio08v391}.

\subsubsection*{Impact of baryons}
In the galaxy formation simulations, Figure~\ref{fig:tng100_radial} shows that haloes in Full-Physics calculations are both rounder (larger $q$ and $s$) and more oblate (smaller $T$) at all radii compared to their DMO counterparts.
This can be clearly seen from the lower attached panels, which show the difference between the median shapes in the Full-Physics and DMO simulations.
We find that the amount of sphericalization depends on the distance to the halo centre: 
the effect of baryons is the largest near the halo centre but reduces towards the virial radius, where differences become negligible.
Taking $10^{12} \msun$ haloes in TNG100, for example, the maximal increase is $\Delta q \approx 0.35$ and $\Delta s \approx 0.25$ near the convergence radius ($r_\text{conv}^{100} = 0.05 R_{200}$). 
The effect of baryons extends far beyond the stellar half-mass radius of the central galaxy, which is typically smaller than 20 per cent of the virial radius.
Even at half the virial radius, there is non-negligible sphericalization of $\Delta q \approx 0.05$ and $\Delta s \approx 0.05$ in all Full-Physics runs.

In general, there is good agreement between the TNG50 and TNG100 results for $10^{11}$ and $10^{12} \msun$ haloes, where the simulation results can be compared in a statistically sound manner.
With a smaller convergence radius, the TNG50 results further demonstrate that the TNG100 results also apply to smaller radii:
the difference between the Full-Physics and DMO results continue to widen down to the TNG50 convergence radius.
There is a small tendency for TNG50 haloes to be more spherical and oblate than TNG100 haloes in the same mass bin.
This is due to the resolution dependency of the galaxy formation model \citep[see e.g. Appendix sections of][]{Pillepich18v473, Pillepich18v475, Pillepich19v490} and is further linked to the relation between galaxy properties and halo shape, which is discussed in Section \ref{sec:variations}.
Briefly, changing the simulation resolution impacts the growth of galaxies (e.g. in terms of star formation and/or strength of baryonic feedback), which subsequently impacts the halo shape.
Further discussion of resolution effects on halo shapes in the Full-Physics runs can also be found in Appendix \ref{sec:resolution}.

\subsubsection*{Comparing TNG100 and Illustris}

On average, halo shapes in TNG100 and the older Illustris simulation are similar.
The biggest discrepancies occur for low-mass and high-mass haloes.
For instance,  $10^{11} \msun$ haloes in  Illustris have larger sphericity (larger $s$) but are more prolate (larger $T$) on average compared to TNG100 haloes.
This is expected since the TNG galaxy formation model introduces changes that decrease the stellar mass function for low- to intermediate-mass galaxies, as well as on the high-mass end of the stellar mass function \cite{Pillepich18v473, Weinberger17v465, Springel17v475}.
This provides some evidence that the halo shapes can indeed be sensitive to the variations in feedback prescriptions in baryonic simulations.


\subsection{Dependence on halo mass}
\label{sec:halo mass}

To examine the dependence of  halo shape on  halo mass, we focus on the local shape at two specific radii: 
(i) the virial radius $R_{200}$ representing the outer halo, and 
(ii) $r_{15} := 0.15 R_{200}$ representing the inner halo, where galaxies are located.
In TNG100-DM, this scale is determined by the smallest radius considered to be converged in low-mass haloes of mass $10^{11} \msun$.
For simplicity, we denote shape parameters measured at $r_{15}$ with a subscript; i.e. $s_{15} := s(r_{15})$, and similarly for $q_{15}$ and $T_{15}$.
Figure~\ref{fig:inner_m200} plots the halo shape parameters as a function of halo mass at the virial radius (top panels) and at $r_{15}$ (lower panels).
Each curve represents the median halo shape, while the shaded regions denote the 25--75th percentile and measure the halo-to-halo variation in the baryonic runs.
The lower attached panels show the difference in the median shapes between the baryonic and DMO counterparts, denoted as $\Delta\med{q} = \med{q_{\rm FP}} - \med{q_{\rm DMO}}$.
Compared to the previous section, we further consider in TNG50 haloes of mass $10^{10} - 10^{11} \msun$.

\subsubsection*{Halo shape at the virial radius and in the inner halo}

At the virial radius (upper panels), there is little difference between the Full-Physics and DMO runs.
Hence, we conclude once again that the impact of baryons is negligible at the virial radius, consistent with the radial profiles shown in Figure~\ref{fig:tng100_radial}.
In general, the parameters $q$ and $s$ exhibit a negative correlation with halo mass, while $T$ exhibits a positive correlation with halo mass: 
more massive haloes tend to be less spherical and more prolate on average.
This is in good agreement with results of previous $N$-body simulations \citep[e.g.][]{Springel04v220,Allgood06v367,Maccio08v391,Schneider2012}.

In the inner halo ($r = r_{15}$), differences between the baryonic and DMO runs become significant.
Although the axis ratios continue to correlate negatively (and positively for $T$) with halo mass, this dependence is no longer monotonic in the Full-Physics runs:
there are clear peaks at $M_{200} \approx 2\times10^{12} \msun$, where haloes tend to be most spherical and oblate.
At this mass, we find that $\med{q_{15}} \approx 0.95$, $\med{s_{15}} \approx 0.75$ and $\med{T_{15}} \approx 0.30$.
From the lower panels, these correspond to a difference of $\Delta \med{q_{15}} \approx 0.25, \Delta \med{s_{15}} \approx 0.2$, and  $\Delta\med{T_{15}} \approx -0.5$ relative to the DMO runs.

Away from the maximum, the impact of baryons in the inner halo decreases. 
In fact, the TNG50 results show that the differences between haloes in the Full-Physics and DMO runs are negligible for $10^{10} \msun$ haloes, where galaxy formation does not appear to have had an impact on halo shapes.

\subsubsection*{Comparison to other hydrodynamic simulations}

\begin{figure*}
    \centering
    \includegraphics[width=\textwidth]{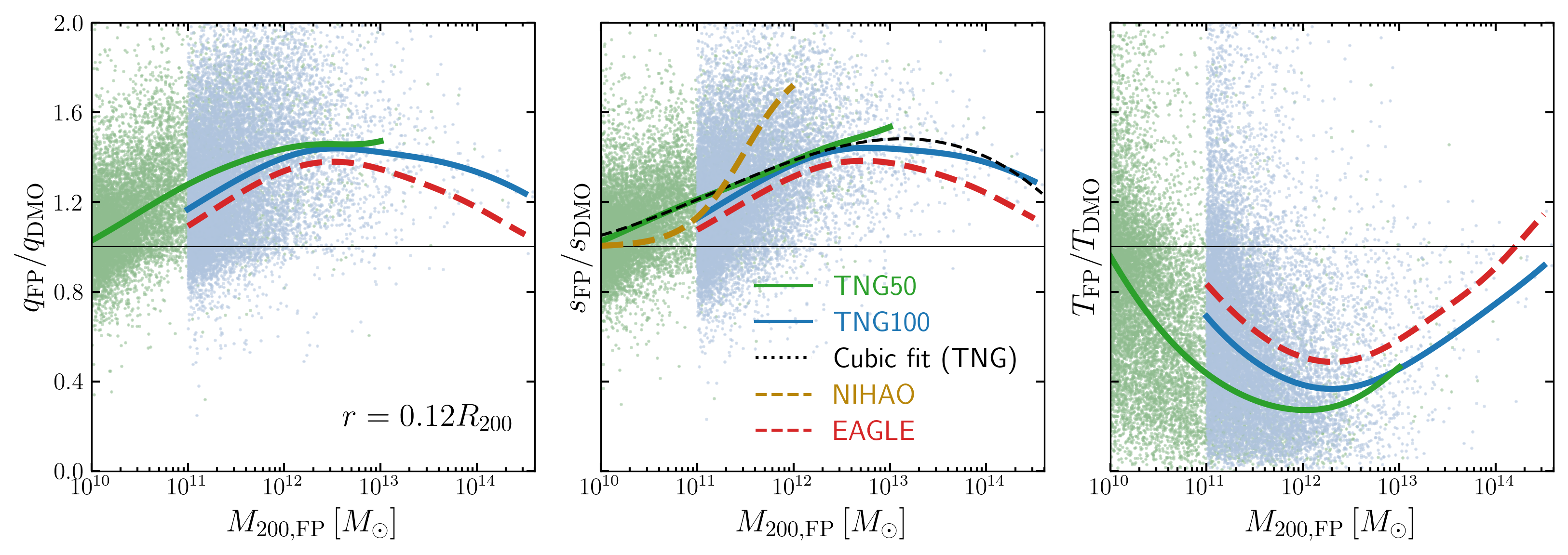}
    \caption{Comparison with NIHAO \protect\cite[][golden dashed lines]{Butsky16v462} and EAGLE \protect\cite[red dashed lines]{EAGLE}.
    We plot the ratio between the inner halo shape of matched haloes between the Full-Physics and DMO simulations, as a function of the halo mass of the Full-Physics counterpart.
    The scatter plots indicate individual matched halo pairs in TNG50 and TNG100, while the median is represented by solid lines.
    For $s_{\rm FP}/s_{\rm DMO}$ (middle plot), a least-squares fit is performed to a cubic equation (Eq.~\ref{eqn:cubic}), and the resulting fit is shown as the black dashed line.
    To facilitate a consistent comparison with NIHAO, here we show results for halo shapes evaluated interior to $r=0.12 R_{200}$ using the reduced shape tensor.
    }
    \label{fig:inner_matched}
\end{figure*}

\begin{table}
    \centering
    \begin{tabular}{c c c c}
        \hline
        $\alpha$ & $\beta$ & $\gamma$ & $\delta$\\
        \hline
        1.50 & 13.09 & -0.127 & -0.026 \\
        \hline
    \end{tabular}
    \caption{Best fit parameters to Eq.~\ref{eqn:cubic} for $s_{\rm FP} /s_{\rm DMO}$, the ratio between the shape parameter $s$ in the baryonic and DMO simulations (middle panel of Figure~\ref{fig:inner_matched}.
    }
    \label{tab:bestfit_nihao}
\end{table}

We compare our TNG results to those from the NIHAO and EAGLE simulations.
NIHAO is a suite of high-resolution simulations of dwarf to MW-size galaxies \citep{Wang15v454}, and we make use of the relation between halo shape and mass presented in Eq.~1 of \cite{Butsky16v462}.
Since the impact of baryons in NIHAO was presented on a halo-to-halo basis, we extract similar results by matching subhaloes between the Full-Physics and DMO runs for TNG50, TNG100 and EAGLE, as described in Section \ref{sec:methods}.
To be consistent with NIHAO, we make the following adjustments for the  shape calculation in this subsection:
(i) halo shapes are calculated using the \emph{reduced shape tensor}, where the contribution of each particle is weighted by $r_\text{ell}^{-2}$, and 
(ii) the inner halo shapes are evaluated at $r=0.12 R_{200}$.

Like TNG, the EAGLE project consists of cosmological hydrodynamical simulations of galaxy formation \citep{EAGLE}.
Although the shapes of dark matter haloes in EAGLE have been analysed and presented in previous work \cite[e.g.][]{Cataldi21v501}, the results presented here are based on our own analysis: 
the halos and subhaloes are identified identically to TNG (i.e. with the same {\sc fof} group finder and {\sc subfind}), 
and halo shapes are calculated with the same code used for Illustris and TNG.
This removes differences resulting from variations of these procedures and ensures that the TNG and EAGLE halo shapes can be compared consistently.

Figure \ref{fig:inner_matched} shows the scatter plots of the relative halo shapes in the MHD and DMO simulations as a function of halo mass.
On the $x$-axis, masses refer to the halo mass of the MHD counterpart. 
The median relations in TNG50 and TNG100 are presented by green and blue curves.
In general, we find that the curves are qualitatively similar to those in the lower attached panels of Figure~\ref{fig:inner_m200}.
Combining haloes from both TNG50 and TNG100, we decided to capture the shape of the ratio $s_{\rm FP}/s_{\rm DMO}$ by fitting a cubic function:
\begin{equation}
f(M) = \frac{s_{\rm FP}}{s_{\rm DMO}} = \alpha + \gamma (\log_{10}M-\beta)^2 + \delta(\log_{10}M-\beta)^3,
\label{eqn:cubic}
\end{equation}
where $M \equiv M_{200}$ is the virial mass of the halo.
By nature of the halo mass function, small haloes strongly outnumber more massive ones; hence we weight each halo  inversely proportional to the halo mass function.
The Levenberg–Marquardt algorithm is applied to solve the least-squares problem, with the best-fit parameters shown in Table~\ref{tab:bestfit_nihao}.
The results show that the maximum average ratio is 1.5, for $\approx 10^{13} \msun$ haloes.
The best-fit curve to the combined TNG50 and TNG100 data is shown by the black dashed line in Figure~\ref{fig:inner_matched}.

The EAGLE results based on our own analysis of the data are plotted in red, which shows good agreement between EAGLE and TNG100.
In particular, the impact of baryons, as captured by the Full-Physics to DMO ratios, exhibits similar dependencies on the halo mass for all three parameters, with maxima/minima located  approximately at $3 \times 10^{12} \msun$.
The values of the ratios in EAGLE are closer to one, signifying a smaller impact of baryons compared to TNG100.
However, we note that this vertical shift is small compared to the halo-to-halo variation, and in particular is less than the difference between TNG50 and TNG100 in the $10^{11} - 10^{12} \msun$ halo mass range.

The NIHAO relation presented  in \cite{Butsky16v462} is shown as the brown dashed curve in the middle panel, which is an increasing $S$-shaped curve.
At a halo mass of $10^{10} \msun$, both the TNG50 and NIHAO results agree that the influence of baryons is negligible.
Although both sets of simulations show that the ratio $s_\text{FP}/s_\text{DMO}$ increases with halo mass up to $\approx 10^{12} \msun$,
NIHAO exhibits a much stronger dependence and the curve increases sharply between $10^{11}$ and $10^{12} \msun$.
At $10^{12} \msun$, the value of the  ratio in NIHAO is ${\approx}$1.7, substantially larger than the best-fit and median curves in TNG and EAGLE.
Since the NIHAO sample only contains haloes hosting dwarf galaxies to MW-size galaxies, the curve does not indicate if the effect of baryons decreases for more massive haloes like in TNG.
The discrepancy between the results from NIHAO and TNG is likely due to the different underlying hydrodynamics solvers and galaxy formation models.

\subsection{Summary}

In this section, we have shown that galaxy formation results in haloes that are more spherical and oblate.
Baryonic physics has the largest impact on halo shapes near the halo centre, and for haloes of mass $2 \times 10^{12} \msun$.
Although comparisons of halo shapes across different simulations can be difficult due to variations in how the halo shape is defined, our comparisons between TNG and other simulations (Illustris, NIHAO and EAGLE) show that important differences are in place.
This can be a result of differences in baryonic feedback implementation and strength, which we quantify in the following section.


\section{Galactic feedback variations}
\label{sec:variations}

In this section, we examine the effects of baryonic-physics variations using the 25~Mpc~$h^{-1}$ smaller boxes. A summary of the nine feedback variations examined has been presented in Table~\ref{table:variations}.

\subsection{Stellar mass -- halo mass relation}

\begin{figure}
	\centering
	\includegraphics[width=0.47\textwidth]{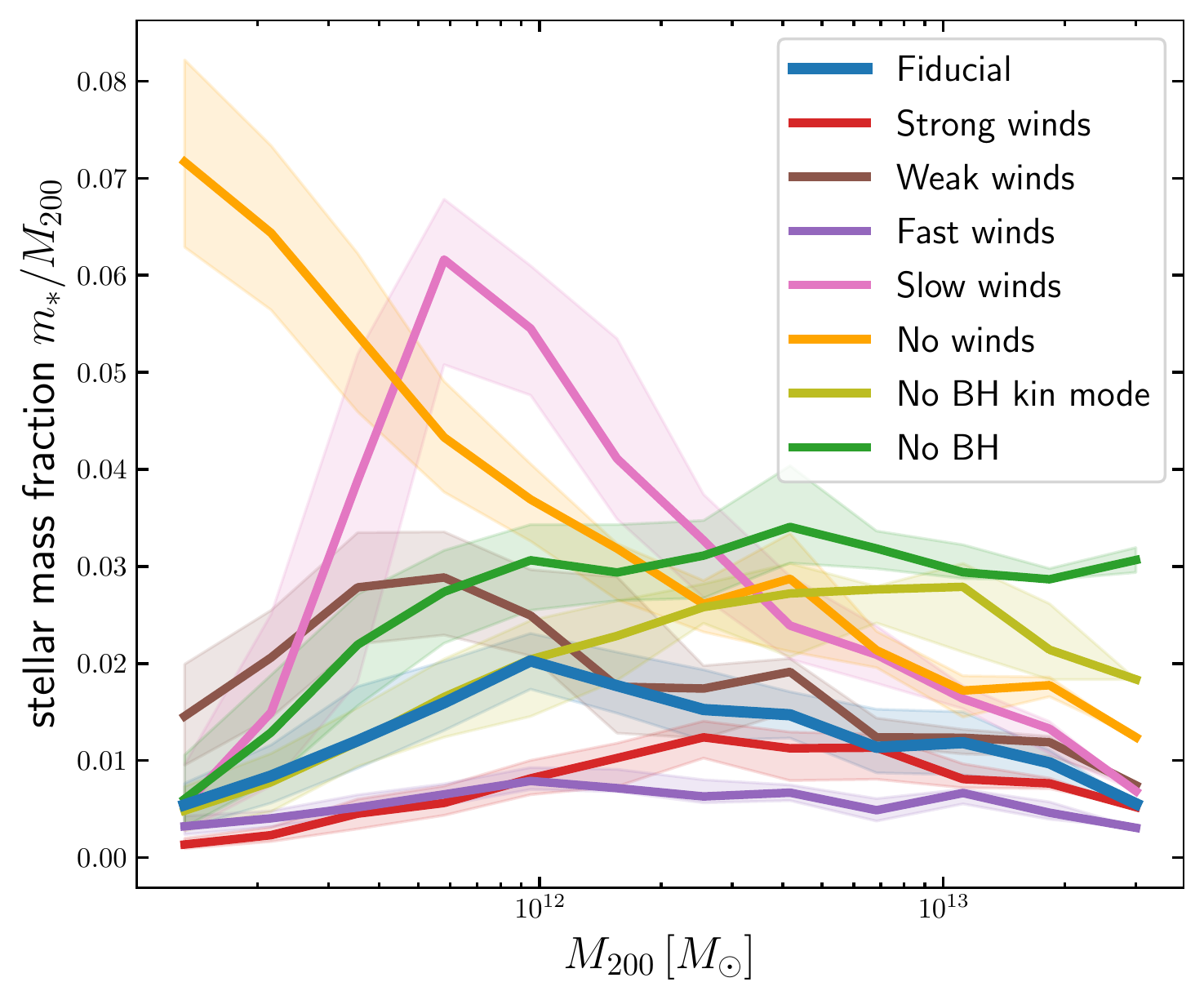}
	\caption{{\bf Stellar mass fraction as a function of halo mass in the small L25n512 boxes.}
	Solid lines denote the median, whereas shaded regions denote the 25th-75th percentiles. Galaxy stellar masses are evaluated within twice their stellar half-mass radius. 
	Reducing the strength of baryonic feedback (winds and BH) leads to an increase in the stellar mass fraction; increasing the strength and speed of galactic winds (\emph{Fast winds} and \emph{Strong winds}) leads to a decrease in the stellar mass fraction.
	}
	\label{fig:smallboxes_galform}
\end{figure}

\begin{figure*}
    \centering
    \includegraphics[width=\textwidth]{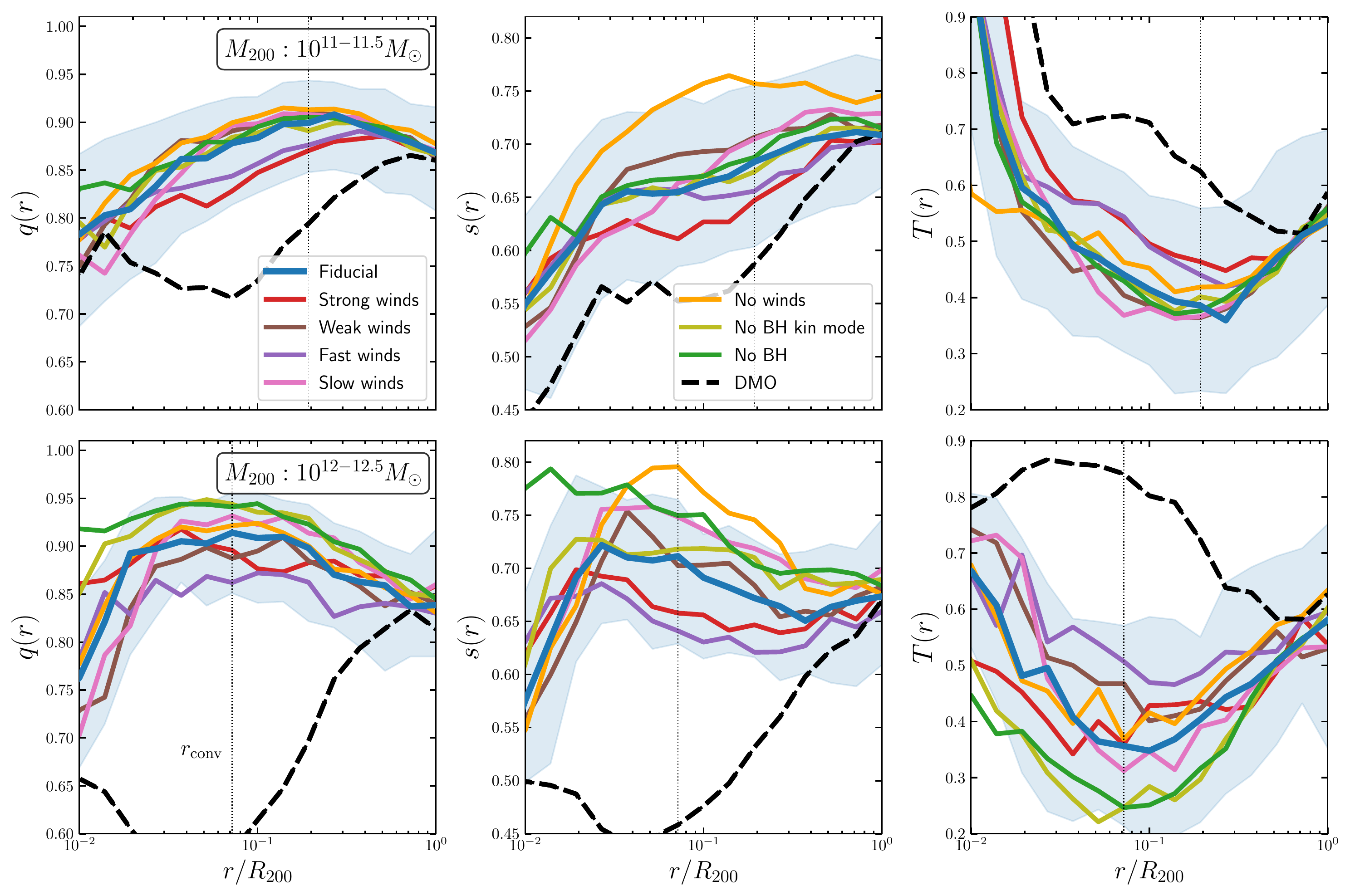}
    \caption{{\bf Effect of changing the feedback prescription on the radial profile of dark matter halo shapes.}
    These simulations are carried out in the small L25n512 boxes.
    The top and bottom rows correspond to haloes of mass $10^{11-11.5} \msun $ and $10^{12-12.5} \msun$, respectively. 
    The solid lines denote the median, while the blue shaded region represents the 25-75th percentile in the fiducial model.
    Changing the feedback prescription has a larger impact on $s$ (middle panels) compared to $q$ (left panels).
    In general, faster and stronger galactic winds result in less spherical (smaller $q$ and $s$) and more prolate (larger $T$) haloes.
    The massive haloes (bottom row) exhibit larger changes in response to variations in feedback prescription.
	An increase in the strength of galactic winds results in less spherical haloes, and conversely for a decrease.
	While black holes only have a minor effect on $10^{11} \msun$ haloes, $10^{12} \msun$ haloes become rounder when black hole feedback is turned off.
	}
    \label{fig:smallboxes_profile}
\end{figure*}

Observations have shown that the galaxy formation efficiency depends non-monotonically on the halo mass and is maximal for Milky Way-sized haloes of $10^{12}\msun$ \citep[e.g.][]{Conroy09v696, Leauthaud12v744}.
We parameterize the galaxy formation efficiency using the stellar mass fraction $m_*/M_{200}$,  where $m_*$ refers to the stellar mass content within twice the stellar half-mass radius\footnote{The galaxy formation efficiency can also be defined as $(m_*/M_{\rm halo}) / (\Omega_b/\Omega_m)$. Also, see \cite{Pillepich18v475} regarding alternative definitions of the stellar mass content.}.

We first show the stellar mass fraction at $z=0$ as a function of halo mass for the various feedback models in Figure~\ref{fig:smallboxes_galform}. 
Analogous curves for the fiducial TNG and Illustris models and flagship runs can be found in \citealt{Pillepich18v473, Pillepich18v475}.
In the fiducial TNG model (blue curve of Figure~\ref{fig:smallboxes_galform}), the stellar mass fraction is maximized at a mass of $10^{12} \msun$, with $m_*/M_{200} \approx 0.02$, about 13\% of the universal baryon fraction ($\Omega_b/\Omega_m$).
For lower mass haloes ($10^{11-12} \msun$), the stellar mass fraction is primarily affected by changes to the galactic winds.
Turning off galactic winds (\emph{No winds}) results in the largest increase in the stellar mass fraction.
Similarly, decreasing the strength and speed of galactic winds both lead to an increase in the stellar mass fraction, evident from the \emph{Slow winds} and \emph{Weak Winds} cases.
Conversely, increasing the wind speed and energy inhibits star formation, and decreases the stellar mass fraction compared to the fiducial model. The corresponding galaxy stellar mass functions at $z=0$ can be found in \citealt{Pillepich18v473}: it is manifest that some of the TNG model-variation runs shown here are known to be ruled out by observational constraints.

Although turning off BH feedback also increases star formation, the resulting increase is less significant compared with changes to the galactic wind, especially at lower halo masses. 
For more massive haloes $(M_{200} >10^{12} \msun)$, BH feedback plays a larger role in determining the stellar mass fraction.
For example, turning off the BH kinetic feedback or turning off BH feedback completely both result in the largest increase to the stellar mass fraction.
Unlike in the smaller haloes, increasing wind speeds appears to be more effective than increasing the wind energy in suppressing star formation.


\subsection{Radial profile of halo shapes with physics variation}

The overall effects of the various feedback models on the profiles of the shape parameters are shown in Figure~\ref{fig:smallboxes_profile}.
For $10^{11-11.5} \msun$ haloes (top row), we find the largest impact by turning off galactic winds (\emph{No winds}), which causes haloes to become significantly rounder compared to the fiducial model.
At $r = 0.15 R_{200}$, the median sphericity is increased by $\Delta \left<s\right> \approx 0.1$ compared to the fiducial model. 
However, such a feedback model is clearly unrealistic and leads to an overproduction of stars in low-mass haloes (Figure~\ref{fig:smallboxes_galform}).
In the \emph{Strong winds} and \emph{Fast winds} cases, both $q$ and $s$ are slightly decreased, while the triaxiality $T$ is increased compared to the fiducial run.
There is little impact from changing the black hole feedback on the halo shapes of low mass haloes.
In the inner halo (at $r=0.15R_{200}$),  the median $\left<q\right>$ varies by $\approx $0.6 across the feedback variations, while the median $\left<s\right>$ varies by $\approx 0.75$ when we ignore the unphysical \emph{No winds} case.

In $10^{12-12.5} \msun$ haloes (bottom row of Figure~\ref{fig:smallboxes_profile}), the feedback variations result in a larger diversity of the median halo shapes.
In the inner halo, decreasing the speed of galactic winds (\emph{Slow winds}) or turning off black hole feedback (\emph{No BH}) result in similar increases to the sphericity $s$: the median $s$ is increased by ${\approx}$0.05.
On the other hand, decreasing the wind energy (\emph{Weak winds}) does not have a noticeable impact on the halo shape profile.
Interestingly, the effect of turning off the black hole kinetic mode (\emph{No BH kin mode}) is similar to turning off BH feedback completely for the parameter $q$, but has no effect on $s$ in the inner halo.
Both increasing the wind speed (\emph{Fast winds}) and decreasing the wind energy (\emph{Strong winds}) result in less spherical haloes, with the largest decrease observed for the \emph{Fast winds} case.

For $10^{12} \msun$ haloes, these changes to the feedback prescription cause the median $\left<q\right>$ and $\left<s\right>$ to vary by ${\approx}0.1$ in the inner halo. Firstly, such variations are comparable to the halo-to-halo variations for the IllustrisTNG fiducial model, indicated as blue shaded areas in Figure~\ref{fig:smallboxes_profile}. Importantly, inner halo shape variations of ${\approx}0.1$ correspond to a change of almost 20 per cent with respect to the DMO halo shape ($s_\text{DMO} \approx 0.55$): different feedback model predictions are closer to one another than to the DMO only ones, also in the case of models whose resulting galaxy populations are known to be unrealistic (e.g. the \emph{No BH} and \emph{No winds} cases).
Referring back to Figure~\ref{fig:inner_matched}, we note that weaker baryonic feedback implementations would help explain why $10^{12} \msun$ haloes in the NIHAO simulations are more spherical than those in the TNG simulations.


\subsection{Dependence on galaxy formation efficiency}
\label{sec:correlation}

\begin{figure*}
	\centering
	\begin{subfigure}{.47\textwidth}
	\includegraphics[width=\textwidth]{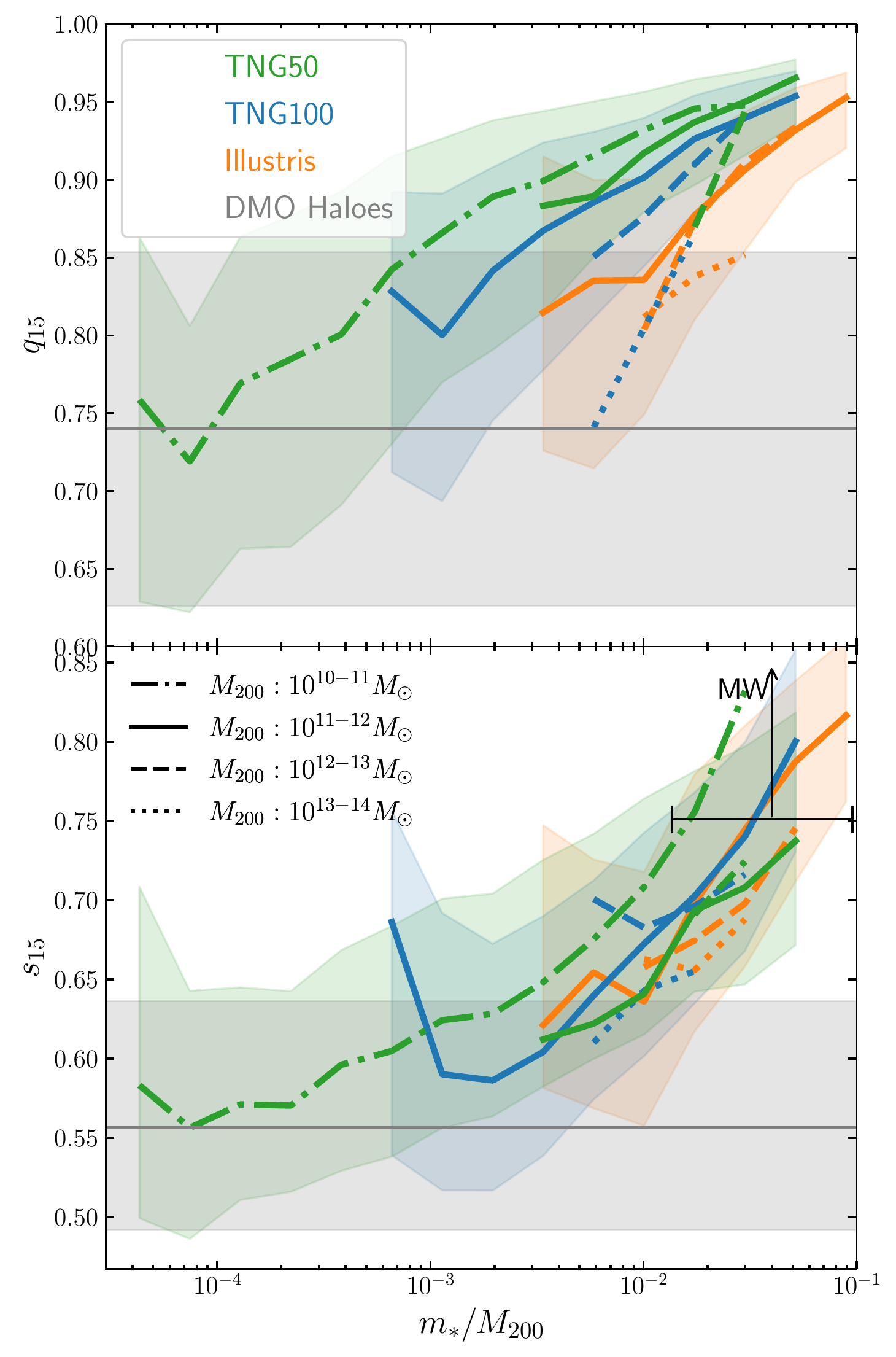}
	\caption{}
	\end{subfigure}
	\begin{subfigure}{.47\textwidth}
	\includegraphics[width=\textwidth]{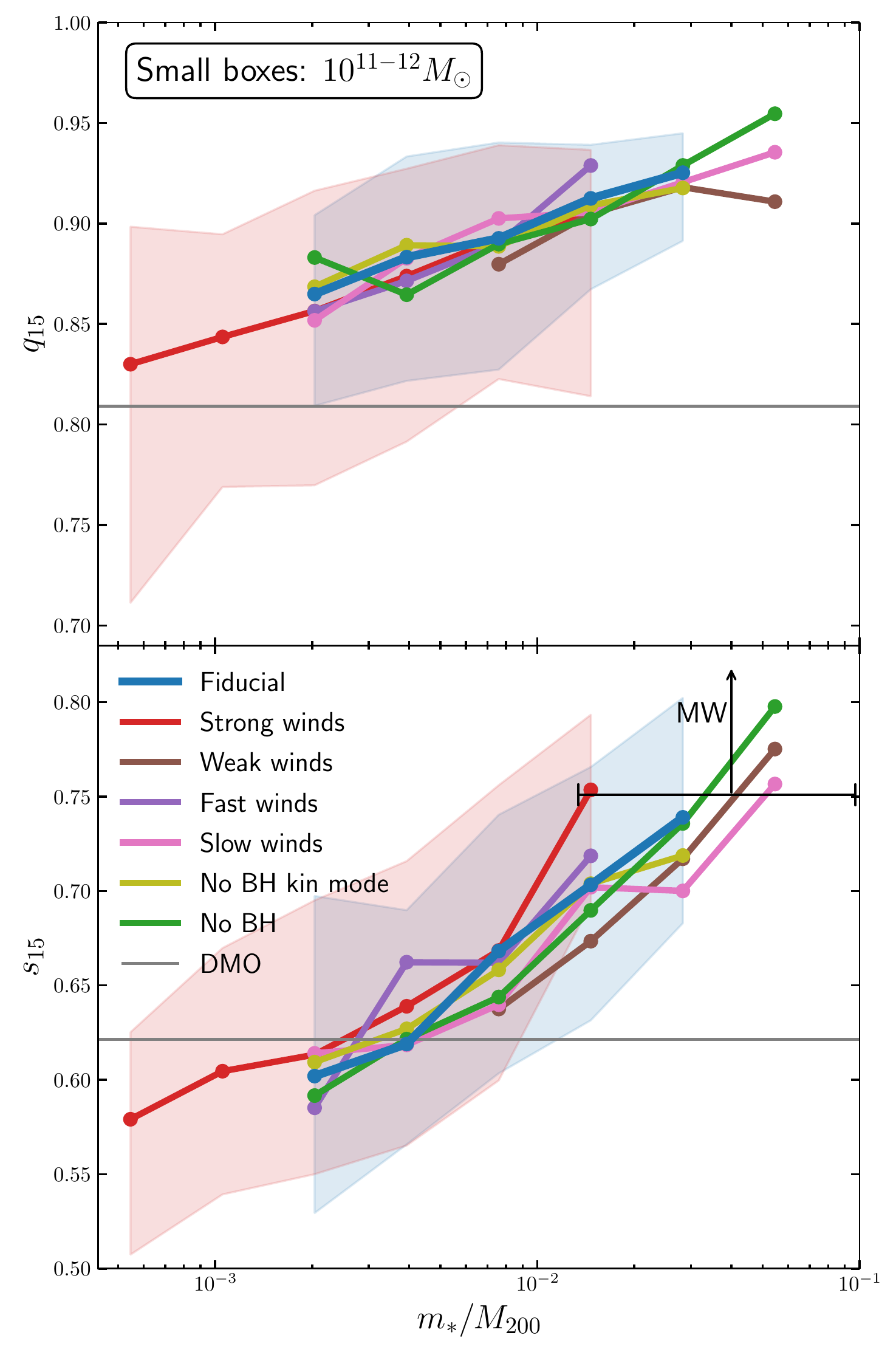}
	\caption{}
	\end{subfigure}
	\caption{{\bf Relationship between the inner halo shape (at $r=0.15 R_{200}$) and the stellar mass fraction.}
	(a) Panels on the left show the shape parameters $s$ and $q$ in the flagship boxes (TNG50, TNG100 and Illustris) for haloes in four mass intervals.
	The curves represent the median, while the shaded regions represent the 25th-75th percentiles.
	The DMO results shown in grey for comparison consist of all resolved haloes with mass $> 10^{11} \msun$ in TNG50 and TNG100.
	The Full-Physics simulations predict a strong dependency of halo shapes on the stellar mass fraction, with maximal sphericalization for haloes with the highest galaxy formation efficiency.
	(b) The panels on the right compares the relationship between inner halo shape and stellar mass fraction for different galaxy formation implementations carried out in the small L25n512 boxes.
	Here, only haloes of mass $10^{11-12} \msun$ are considered.
    The different curves are in general agreement, which suggests that halo shape differences caused by varying baryonic feedback in the simulations can be explained by its cumulative effects on the stellar mass fraction.
    A lower bound for the sphericity $s$ of the Galaxy (MW) from observations is shown for comparison, with the horizontal error bar representing the uncertainty in the stellar mass fraction.
	}
	\label{fig:smallboxes_corr}
\end{figure*}

\begin{table*}
	\centering
	\begin{tabular*}{0.6\textwidth}{@{\extracolsep{\fill}}l c c}
		\midrule
		Simulation      & Pearson $r$ (p-value) & Spearman $\rho$ (p-value) \\
		\hline
		1. Fiducial        & 0.259 $(4.9 \times 10^{-17})$  & 0.343   $(1.7 \times 10^{-17})$\\ [1ex]
		                
        2. Strong winds    & 0.338 $(2.3 \times 10^{-10})$  & 0.259   $(2.3 \times 10^{-10})$\\ [1ex]          

        3. Weak winds      & 0.282 $(2.3 \times 10^{-12})$  & 0.276   $(5.9 \times 10^{-12})$\\ [1ex]       

        4. Fast winds      & 0.223 $(1.2 \times 10^{-7})$  & 0.214   $(3.6 \times 10^{-7})$\\ [1ex]       

        5. Slow winds      & 0.444 $(2.8 \times 10^{-30})$  & 0.455   $(4.7 \times 10^{-32})$\\ [1ex]      

        6. No BH kinetic model & 0.365 $(1.0 \times 10^{-19})$  & 0.373   $(1.3 \times 10^{-20})$\\ [1ex]

        7. No BH           & 0.435 $(9.4 \times 10^{-29})$  & 0.440   $(1.8 \times 10^{-29})$\\ [1ex]      

		8. TNG100          & 0.352 $(0)$  & 0.368   $(0)$\\ [1ex]       

		9. TNG50           & 0.301 $(0)$  & 0.355   $(0)$\\ [1ex]       
		\bottomrule
	\end{tabular*}
	\caption{
	Pearson and Spearman correlation coefficients and their associated p-values between the shape parameter $s$ and the stellar mass fraction.
	The first seven rows are results for model variations in the small boxes, while the last two rows correspond to the larger boxes, TNG100 and TNG50.
	For TNG100 and the small boxes, all haloes of mass greater than $10^{11} \msun$ are considered.
	For TNG50, all haloes of mass greater than $10^{10} \msun$ are considered.
	All coefficients are positive and significant, pointing to a strong correlation between the sphericity of a halo and its galaxy formation efficiency.
	}
	\label{table:smallboxes_corr}
\end{table*}

The differences in halo shape profiles induced by the model variations suggest that the galaxy formation efficiency is closely linked to 
the resulting halo shapes in baryonic simulations, since the halo shape reflects the competition between the competing processes of star formation and feedback.
The link between halo shape and the growth of the baryonic component is further supported by  controlled $N$-body experiments that study the effect of baryon condensation by simulating the growth of baryonic disks within triaxial haloes.
For example, \cite{Debattista08v681} found that the final halo shape depends on the mass of the baryonic disk, and \cite{Kazantzidis10v720} concluded that the halo response depends most strongly on the overall gravitational importance of the disk.

To verify these conclusions, we plot the shape parameters $q$ and $s$  in the inner halo ($r = 0.15 R_{200}$) as a function of the stellar mass fraction in Figure \ref{fig:smallboxes_corr}.
The left panels show the results in the large boxes (TNG50, TNG100 and Illustris), with curves denoting the median of the distributions for haloes of different masses.
In general, the shape parameters correlate positively with the stellar mass fraction: 
haloes are rounder (larger $q$ and $s$) and more oblate (smaller $T$) with increasing stellar mass fraction.
At a stellar mass fraction of $m_*/M_{200} \gtrsim 0.05$, the average sphericity is  $\left<s\right> \gtrsim 0.75$, significantly larger than the average DMO value (horizontal grey line and shaded region). 
With decreasing stellar mass fraction, halo shapes approach the DMO value.
From the TNG50 results, we find that sphericalization of the dark matter halo becomes negligible for $m_*/M_{200} < 10^{-4}$.
Note that almost most curves exhibit an upturn in the halo shapes at the lower end of the stellar mass fractions, which we believe is an indirect resolution effect.

The results here also suggest that even though the stellar mass fraction normalizes for the halo mass, there is residual mass dependence of the halo shape -- stellar mass fraction relation: 
at a fixed stellar mass fraction, less massive haloes tend to be more spherical than their more massive counterparts.
This is associated with the DMO halo shape -- halo mass dependence previously touched upon in Section~\ref{sec:halo mass}, which continues to hold in baryonic simulations after controlling for the galaxy stellar mass fraction.

The results of the physics variations shown in the right panels of Figure \ref{fig:smallboxes_corr} paint a similar picture, for haloes of $10^{11-12} \msun$ only:
the median curve for each variation displays a positive correlation between halo shape and stellar mass fraction.
In particular, there is substantial overlap across all curves, and the differences across the cases are small ($\lesssim 0.25$) at fixed stellar mass fractions.
This provides a strong indication that the shape of the average halo in all baryonic simulations is largely determined by its  galaxy formation efficiency, regardless of the exact galaxy formation model used for the simulation. At halo masses of $10^{11-12} \msun$, the outcome of the model-variation runs also differs substantially from the DMO predictions, for most implementations and stellar mass fractions larger than $1-2$ per cent.

For MW-sized haloes, the most spherical haloes in TNG have sphericities of $s_{15} \approx 0.85$.
This is consistent with the simulations of \cite{Abadi10v407}, which considered the maximal effects of baryons by neglecting stellar formation and other forms of baryonic feedback.
In their analysis, they found that this model resulted in a halo with sphericity $s \sim 0.85$ approximately constant across radii.


\subsection{Comparison to observations of the Milky Way halo shape}

\begin{figure*}
    \centering
    \includegraphics[width=0.9\textwidth]{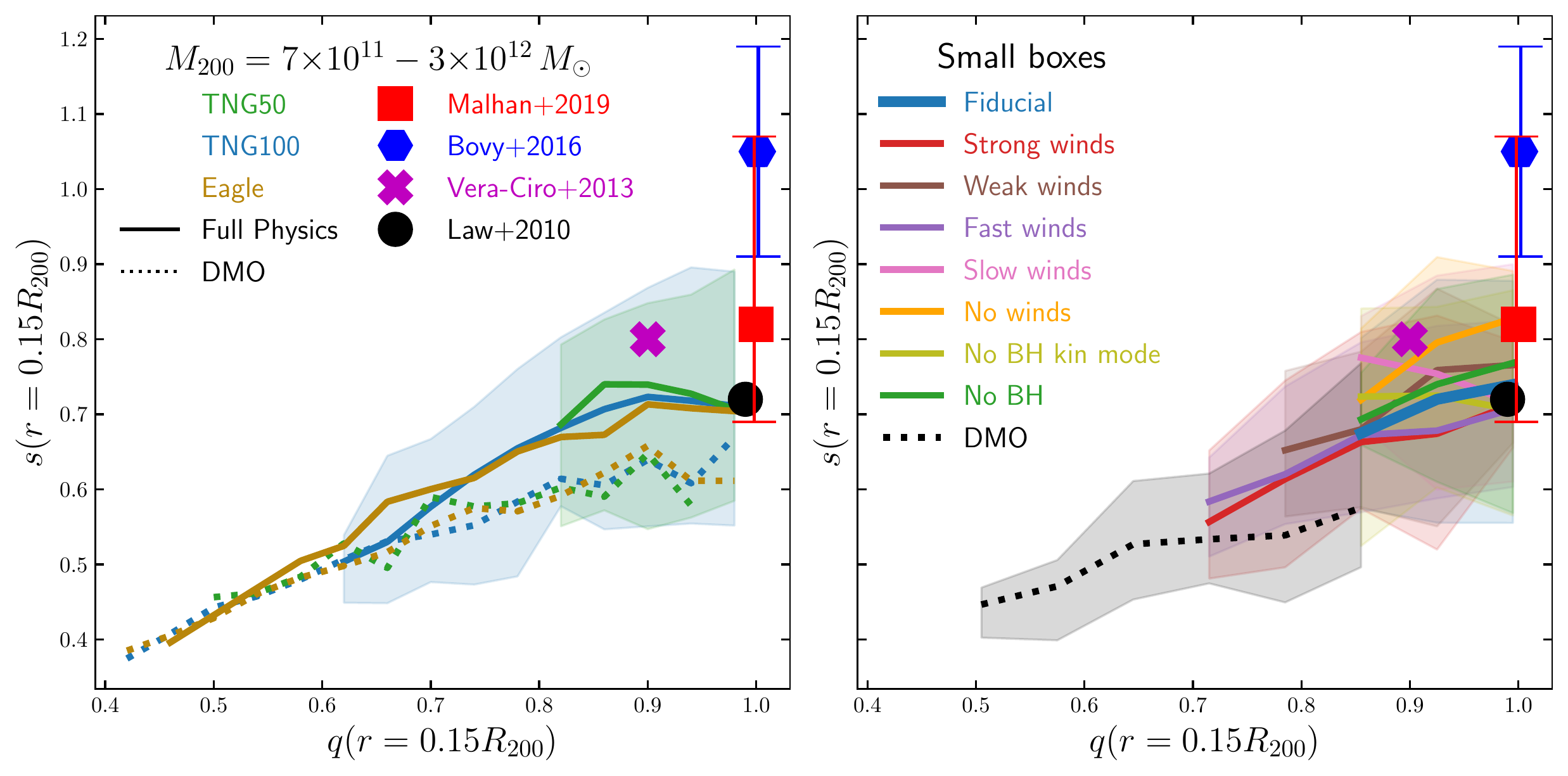}
    \caption{{\bf Distribution of the sphericity $s$ as a function of the parameter $q$ in the inner halo,} for haloes of mass $M_{200}=7\times10^{11}-3\times10^{12} \msun$.
    The left panel shows the results from TNG50, TNG100 and EAGLE, while the right panel shows the results for the different baryonic feedback models in the smaller boxes.
    Solid lines represent results from MHD simulations, while dashed lines represent results from the DMO simulation.
    Here, the shaded region represents the $1-\sigma$ interval in the MHD simulations.
    The red, blue, magenta and black symbols denote observational estimates of the MW halo shape by \protect\cite{Malhan19v487}, \protect\cite{Bovy16v833}, \protect\cite{Vera-Ciro13v773} and \protect\cite{Law10v714}, with error bars denoting the $1-\sigma$ uncertainty (if available).
    }
    \label{fig:smallboxes_QS}
\end{figure*}

\begin{figure*}
    \centering
    \includegraphics[width=0.8\textwidth]{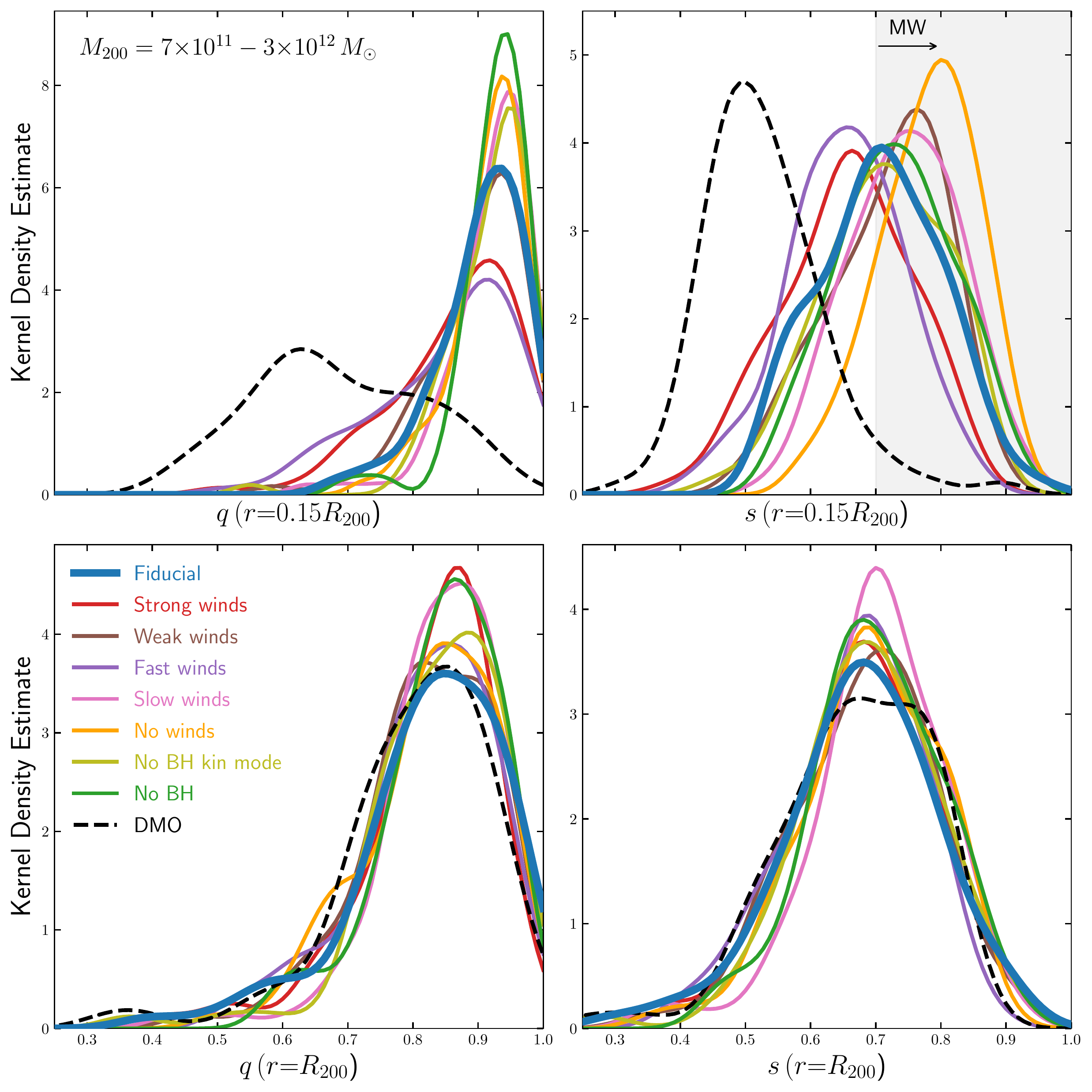}
    \caption{{\bf Effect of physics variations on the distributions of shapes of Milky Way-sized dark matter haloes  ($M_{200}=7\times10^{11}-3\times10^{12} \msun$) in the smaller boxes.}
    The distribution for each feedback model is represented by a kernel density estimate. 
    The grey shaded region denotes the range of observational estimates of the MW sphericity $s$ in the inner halo.
    In the inner halo ($r=0.15R_{200}$, top row), all the feedback models result in haloes that are substantially more spherical than those of the DMO simulation. 
    Substantial overlap remains across the feedback variants, rendering it difficult to distinguish between these models based solely on measurements of the MW halo shape.
    At the virial radius (bottom row), regardless of the feedback model, the effect of baryons becomes negligible. 
    }
    \label{fig:smallboxes_MW}
\end{figure*}

In this paper, we have found that baryons result in rounder haloes for a large fraction of the virial radius.
In addition, we demonstrated that the amount of sphericalization correlates strongly with the stellar mass fraction of the halo, confirming the conclusions  of the controlled $N$-body experiments conducted by \cite{Debattista08v681} and \cite{Kazantzidis10v720}.
Our analysis is also in agreement with the findings of previous hydrodynamic simulations, which had found that the change in halo shape depends on aspects of the baryonic physics; e.g. gas cooling, star formation and galactic feedback \citep[e.g.][]{Bryan13v429,Butsky16v462,Prada19v490}.
By considering a variety of galactic feedback models, we showed that the correlation between halo shape and stellar mass fraction is consistent even when baryonic physics are modified in the simulations.

In our Galaxy, the shape of the inner halo (${\lesssim} 60$~kpc) has been inferred through observations of the motion of individual stars.
Kinematics of halo stars combined with equilibrium modelling using the Jean's equations can be used to infer force fields and thus the MW halo shape \cite[e.g.][]{Loebman2012,Bowden16v460}.
Stellar streams formed from the tidal stripping of satellite galaxies or globular clusters can also be used to constrain the halo potential.
Examples of stellar streams that have been used include that of the Sagittarius dwarf galaxy \citep{Ibata01v551,Law10v714,Vera-Ciro13v773}, as well as the Pal 5 and GD-1 tidal streams \citep{Bovy16v833}.
In general, these observations indicate that the MW inner halo is substantially more spherical than DMO predictions:
$s$~=~0.72 \citep[20-60 kpc][]{Law10v714}, $s$~=~0.8 \citep{Vera-Ciro13v773},  $s=$~1.05 \citep[20~kpc][]{Bovy16v833} and
$s=$~0.82 \citep{Malhan19v487}, at radii somewhat smaller than $0.2-0.3$ times the virial radius of the Galaxy.


In Figure~\ref{fig:smallboxes_corr}, we have presented a quick comparison to the MW observations by indicating a lower bound for the stellar stream measurements of the sphericity ($s \gtrsim 0.75$).
The horizontal error bar represents the uncertainty in the stellar mass fraction of our Galaxy, ranging approximately between 0.01 and 0.1, consistent with  observational estimates \cite[e.g.][]{Zaritsky20v888}.
The stellar mass fraction the Galaxy corresponds to is the higher end in the simulations, where the observational estimates and numerical predictions are in good agreement.

A more thorough comparison between observational and simulation results is presented in Figure~\ref{fig:smallboxes_QS}, which plots the sphericity $s_{15}$ as a function of $q_{15}$ for MW-size haloes ($7\times 10^{11} - 3\times 10^{12} \msun$).
The left panel shows the results in TNG50 and TNG100 as well as EAGLE, while the right panel shows the results for the model variations.
Estimates of the MW inner halo shape from stellar streams are shown for comparison, with error bars denoting the 1-$\sigma$ error (if provided by the reference).
Three of the observations (\citealt{Vera-Ciro13v773}, \citealt{Bovy16v833} and \citealt{Malhan19v487}) indicate that the MW halo is axisymmetric in the disk plane, hence $q \equiv b/a$ is almost exactly one.

In general, the shift to larger sphericities in the MHD; i.e. Full-Physics, runs means that TNG50 and TNG100 are in much better agreement with observations compared to their DMO counterparts.
Since \cite{Bovy16v833} estimates the MW halo to be almost completely spherical, we find a slight disagreement with their result at the $1-\sigma$ level.
Interestingly, this disagreement is not reduced even when we consider weaker feedback formulations, shown in the right panel.
Although the median $s$ increases for models with weaker feedback (e.g. \emph{Weak winds} and \emph{No BH} cases), the upper percentiles of the distributions (in $s$) do not shift upwards sufficiently to meet the \cite{Bovy16v833} inferences: in fact, the galaxies produced by those models are unrealistic and so a possible agreement in MW halo shapes would still point towards a tension.

We illustrate these findings more clearly in Figure~\ref{fig:smallboxes_MW}, which plots the kernel density estimates (KDE) of the distributions of $q$ and $s$ in the small boxes.
The upper and lower panels show the halo shape distributions in the inner halo and at the virial radius, respectively.
Although decreasing the baryonic feedback strength tends to shift the distribution of $s_{15}$ (upper right) to larger values of $s$, the biggest changes are for haloes with $s_{15} \lesssim 0.85$. 
If accurate, the \cite{Bovy16v833} results would rule out the DMO, \emph{Strong winds} and \emph{Fast winds} models, which have negligible fractions of haloes with $s_{15} > 0.9$.
In the remaining models, the fractions of haloes with $s_{15} > 0.9$ are small but non-zero, thus it is possible for these models to produce halos compatible with the \cite{Bovy16v833} results.
Since observations generally assume or find that the minor axis of the halo and of the stars (i.e. galaxy) coincide, we note that it maybe be possible to improve the agreement with observations by projecting the halo shape in the simulation perpendicular to the stellar axis \citep{Chua19v484}.

In general, the results shown in this paper indicate that the inner dark-matter halo shapes predicted by Full-Physics models differ substantially from those from DMO calculations and that, even for unrealistic feedback models, their predictions are closer to one another than to the DMO results. However, Figure~\ref{fig:smallboxes_MW} points to a difficulty in using dark matter halo shape as a constraint on baryonic feedback models with individual observational data points:
changing the baryonic physics results in distributional shifts that are small compared to the large halo-to-halo variation ($\sigma \approx 0.15$) in the halo.
This means that observations have to contain a large sample of haloes to distinguish between the various models.
Unfortunately, inferring the 3-dimensional shapes of dark matter haloes is challenging since current methods rely strongly on stellar observations and is thus only feasible for nearby galaxies.

In these cases, none of the feedback variations affects the halo shape far from the halo centre.
This is illustrated in the lower panels of Figure~\ref{fig:smallboxes_MW}, which show that the distributions of $q$ and $s$ remain close to that of the DMO run.

\section{Summary}
\label{sec:conclusions}

We have used a suite of cosmological simulations to investigate the impact of galaxy assembly on the shape of dark matter haloes at redshift $z=0$.
To elucidate the effects of the fiducial galaxy formation model, we have examined haloes from TNG50 and TNG100, which are both part of the suite of magneto-hydrodynamic cosmological simulations IllustrisTNG.
Due to the high resolution of TNG50, we were able to reliably resolve halo shapes down to halo masses of $10^{10} \msun$.
In total, we have analysed  statistically significant samples of ${\approx}10000$ and ${\approx}14000$ haloes in TNG50 and TNG100, respectively, spanning a total halo mass range of $10^{10}$ to $10^{14} \msun$.
Additionally, we have also investigated a set of 25$h^{-1}$~Mpc smaller boxes, which have numerical resolutions comparable to that of TNG100.
A total of nine model variations have been analysed, including five galactic wind feedback variations and two black hole feedback variations: these have allowed us to bracket the effects on simulated halo shapes due to lingering uncertainties in the galaxy formation models, also including unrealistic feedback implementations.

Using an iterative algorithm for the unweighted shape tensor, we have quantified halo shapes in ellipsoidal shells as a function of radius, which are summarized with the parameters $q(r) \equiv b/a$, $s(r) \equiv c/a$ and the triaxiality parameter $T(r) \equiv (1-q^2)/(1-s^2)$.
When focusing on the inner halo, we have chosen to represent the inner halo shape by the measurement at $r_{15} \equiv 0.15R_{200}$, where haloes across the examined mass range are determined to be resolved.
We compared the TNG haloes to their counterparts in the dark matter-only analogues TNG50-DM and TNG100-DM, as well as the previous generation Illustris.
Furthermore, we have also compared our results to previous simulations and observational estimates of the MW halo shape.
We summarize our results as follows:

\begin{itemize}

\item 
The convergence radii in TNG100-DM and TNG50-DM are determined to be $r_\text{conv}$ = 10~kpc and 4~kpc, respectively.
These values correspond to 15 per cent of the virial radius for $10^{11} \msun$ haloes in TNG100, and 10 per cent of the virial radius for $10^{10} \msun$ haloes in TNG50.\\

\item 
In TNG50 and TNG100, galaxy formation results in haloes that are more spherical and oblate on average than their DMO counterparts (Figure~\ref{fig:tng100_radial}).
The sphericalization extends almost through the entire halo: the difference between the median halo shapes in TNG and the DMO runs is the largest near the halo centre and only becomes negligible near the virial radius.
For haloes of mass $10^{12} \msun$, the median $\left<q\right>$ and $\left< s \right>$ exceeds that of the DMO runs by almost 0.5 and 0.3 at $r = 0.02 R_{200}$.
We find good agreement between the results of TNG50 and TNG100, although some minor discrepancies can be noticed due to the dependency of the baryonic physics models on simulation resolution.\\

\item 
Considering the inner halo shape measured at $r=0.15R_{200}$, we demonstrate that the shape parameters $q$, $s$ and $T$ no longer follow the monotonic trend with halo mass that previous DMO simulations have found.
The results from TNG50 and TNG100 predict that haloes near a mass of $M_{200} = 2 \times 10^{12}$ are the roundest and most oblate (Figure~\ref{fig:inner_m200}).
In contrast, the current and previous DMO simulations predict that smaller haloes are more spherical,  i.e. $q$ and $s$ anti-correlate with the halo mass.
For small haloes of mass $10^{10} \msun$, the effect of baryons is negligible.\\

\item 
Based on our own post-processing and analysis of the EAGLE data, we find good agreement between TNG100 and EAGLE (Figures \ref{fig:inner_matched} and \ref{fig:smallboxes_QS}).
We highlight that the impact of baryons, as captured by the Full-Physics to DMO ratio, has a similar dependence on halo mass in both simulations (Figure~\ref{fig:inner_matched}). 
\\

\item 
Through the feedback variation runs in smaller simulation boxes (Table~\ref{table:variations}), we show, on the wake of the results of \citealt{Weinberg08v678} and \citealt{Pillepich18v473}, that the galaxy formation efficiency as parameterized  by the stellar mass fraction is strongly affected by the changing feedback strength.
In general, increasing (decreasing) the galactic wind speed or energy suppresses (increases) the stellar mass fraction.
Turning off black hole feedback has the effect of increasing the stellar mass fraction, primarily at larger halo masses.\\

\item  
We show that varying the feedback prescription has an impact on the median halo shape parameters (Figure~\ref{fig:smallboxes_profile}):
on average, stronger feedback prescriptions (increasing the galactic wind energy or speed) causes haloes to become less spherical and more prolate.
Conversely, weaker feedback prescriptions (through the reduction of galactic wind speeds or suppression of black hole feedback) result in more spherical and oblate haloes.
However, in  the inner haloes, we find a maximum change in the median parameters $\left<q\right>, \left<s\right>, \left<T\right>$ of around 0.1 across the nine feedback-model variations: this is comparable to the halo-to-halo variation for the fiducial IllustrisTNG model, which has standard deviation $\sigma \approx 0.15$, and makes the results of any feedback model studied here closer to one another than to the DMO expectations.\\

\item 
We demonstrate that a positive correlation is in place between the shape of the inner halo and the stellar mass fraction: haloes with larger stellar mass fractions are rounder and more oblate (Figure~\ref{fig:smallboxes_corr}).
At a fixed stellar mass fraction, the predicted inner halo shapes are consistent across the simulations examined (TNG50, TNG100 and the feedback variations), within average variations of 0.1. 
In particular, we note that the effect of baryons is negligible when the stellar mass fraction is $m_*/M_{200} \lesssim 10^{-4}$. However, for stellar mass fractions larger than about $1-2$ per cent, all feedback variation runs predict average halo shapes that are substantially different from those from DMO calculations and more compatible with one another than with the DMO results, even in the case of manifestly unrealistic feedback implementations. \\

\item  
Comparing with several observational estimates of the MW halo shape, we find good agreement between observations and the fiducial model predictions from TNG50 and TNG100 at the 1--$\sigma$ level (Figure~\ref{fig:smallboxes_QS}).
With the small boxes, we demonstrate that the eight baryonic models (excluding the DMO model) are consistent with a value of $s_\text{MW} = 0.7-0.8$ for the MW halo shape, due to the large overlap in the parameter distributions (Figure~\ref{fig:smallboxes_MW}).
Although these values disfavour the models with stronger galactic feedback, the halo shape alone does not provide a very strong constraint on baryonic feedback models, but may do so in combination with complementary galaxy diagnostics and statistics.

\end{itemize}

\section*{Acknowledgements}

MV acknowledges support through NASA ATP grants 16-ATP16-0167, 19-ATP19-0019, 19-ATP19-0020, 19-ATP19-0167, and NSF grants AST-1814053, AST-1814259,  AST-1909831 and AST-2007355.

\section*{Data Availability}

The snapshot data for TNG50 and TNG100 can be accessed at \url{www.tng-project.org} and for Illustris at \url{www.illustris-project.org}. The corresponding data-release papers are \citealt{Nelson19TNGrelease} and \citealt{Nelson15v13}.
Other data used in this paper will be shared upon reasonable request to the corresponding author.

\bibliographystyle{mnras}
\bibliography{references}

\appendix
\section{Resolution convergence}
\label{sec:resolution}

\begin{figure*}
    \centering
    \includegraphics[width=0.86\textwidth]{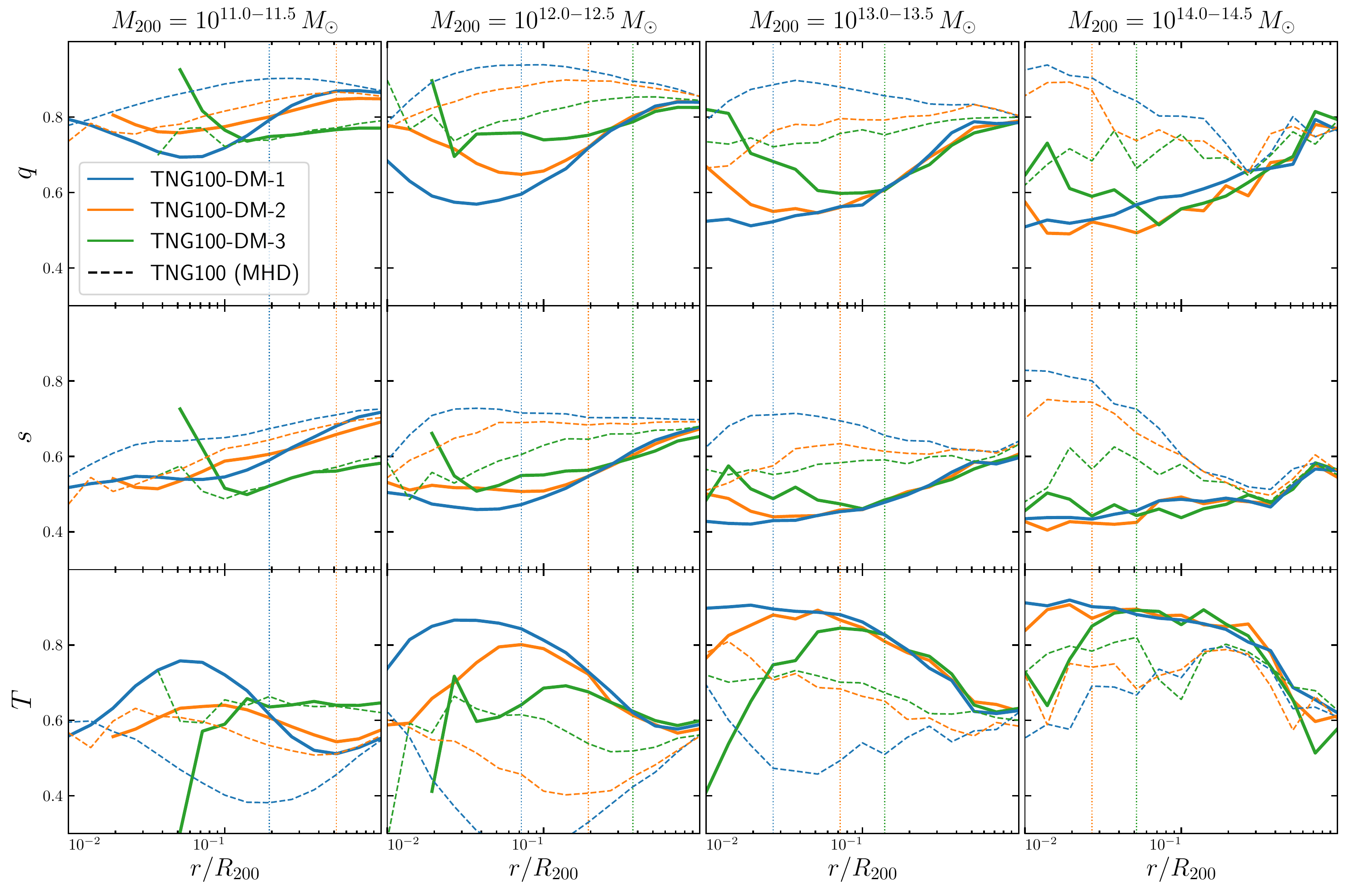}
    \caption{Effect of resolution on the median dark matter halo shapes in TNG100-DM (solid lines) and TNG100 (MHD, dashed lines) in various mass bins.
    Dotted lines show the median convergence radii $r_{\rm conv}$ as derived from $\kappa(r) = 7$ (Eq.~\ref{eqn:power}).
    For $r > r_\text{conv}$, the median DMO halo shapes are converged; i.e. the lower resolution runs match the highest resolution TNG100-DM-1.
    The halo shapes of $10^{11}\msun$ haloes (left panels) are not converged at all in the lowest resolution TNG100-DM-3.
    In the MHD runs (dashed lines), halo shapes do not appear to converge with resolution, due to the additional resolution dependence of the galaxy formation physics.
    }
    \label{fig:resolution-tng100}
\end{figure*}

\begin{figure*}
    \centering
    \includegraphics[width=0.7\textwidth]{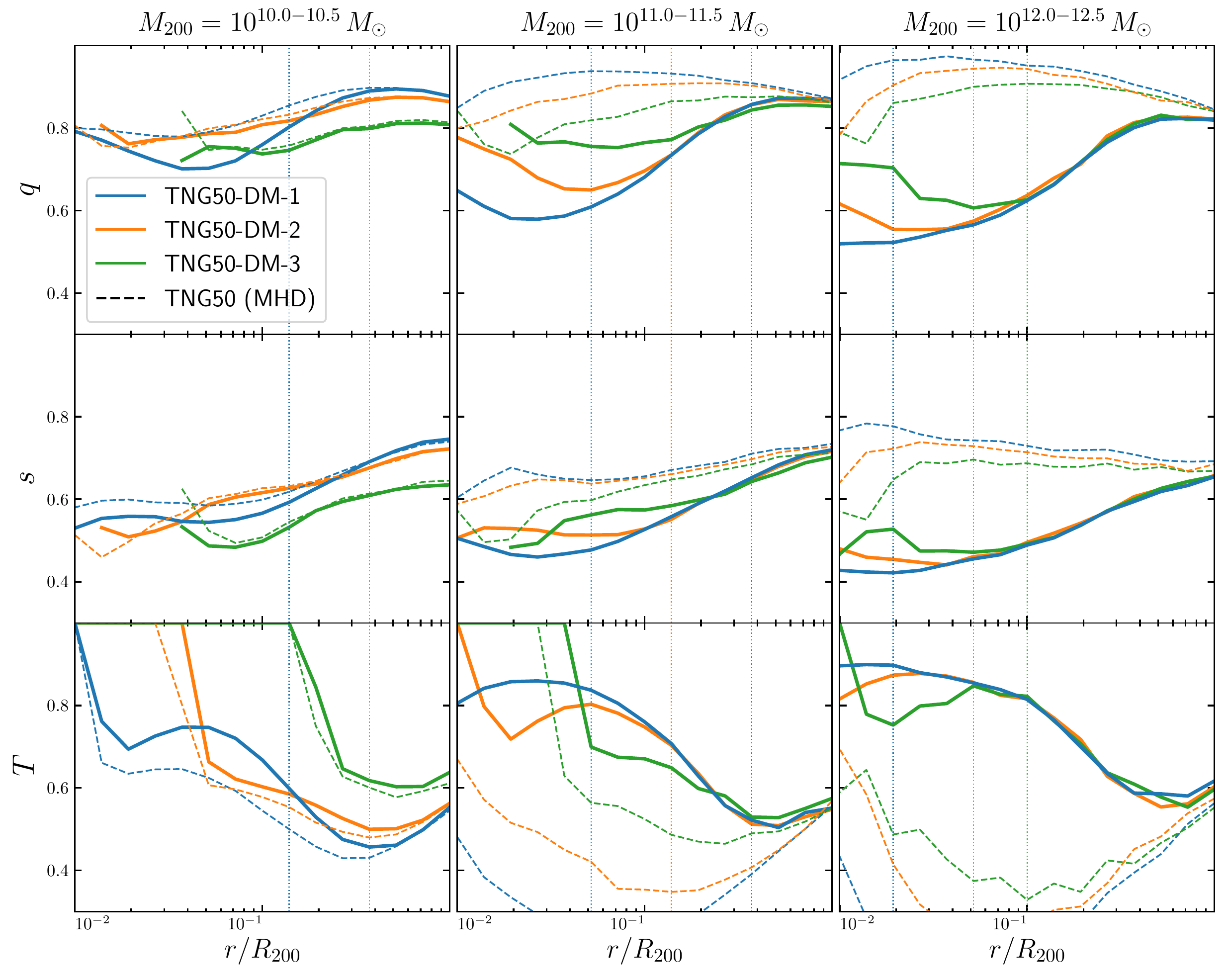}
    \caption{Effect of resolution on the median dark matter halo shapes in TNG50-DM (solid lines) and TNG50 (MHD, dashed lines), for haloes between $10^{10}$ and $10^{12.5} \msun$.
    In particular, note that $r=0.15 R_{200}$ is resolved in $10^{10} \msun$ haloes for the highest resolution run.
    }
    \label{fig:resolution-tng50}
\end{figure*}

In this work, we have primarily considered the TNG100 runs with the highest resolution, TNG100-1 and TNG100-DM-1.
Here, we test the numerical convergence of the median halo shape profiles, presenting in Fig. \ref{fig:resolution-tng100} the results for three different resolution levels, and haloes in three different mass bins.
Each lower resolution level is coarsened by a factor of 8 in mass resolution and a factor of 2 in the softening length.
The dotted vertical lines show the median convergence radii $r_{\rm conv}$ as derived from Eq.~\ref{eqn:power} across the haloes in the mass bin.
We find in the DMO case (solid lines) good convergence of the median halo shape for $r > r_{\rm conv}$.
Due to the low resolution of TNG100-DM-3, the halo shapes of $10^{11}\msun$ haloes (left panels) are not converged at any radii.

For comparison, we plot also results from TNG100 (dashed curves).
In these MHD runs, the halo shapes are not converged with resolution, even in the region  $r > r_{\rm conv}$.
Instead, haloes from the higher resolution runs are more spherical, due to the resolution dependence of the baryonic physics.
In the TNG galaxy formation model, lowering the  simulation resolution results in lower galaxy formation efficiencies \citep{Pillepich18v475}, causing haloes to be less spherical as well.

Compared to circular velocities, the convergence of halo shapes is more demanding, with a larger minimum radius $r_{\rm conv}$ of the converged region.
Although $\kappa >7$ describes well the converged regions of halo shapes in DMO simulations, the situation is more complicated in hydrodynamic simulations since the baryonic physics implementations can also depend on resolution.
For example, the TNG galaxy formation model when applied to a larger box of side-length 300~Mpc results in the systematically lower stellar masses of galaxies compared to the smaller box TNG100 \citep{Pillepich18v475}, due to the lower mass and spatial resolution in TNG300.
For this reason, we use Eq.~\ref{eqn:power} applied to the DMO runs to identify the regions where the effects of two-body relaxation can be neglected.

\end{document}